\documentclass[twocolumn]{aastex7}

\usepackage{enumitem}
\usepackage{amsmath}
\usepackage{latexsym}
\usepackage{graphicx}
\usepackage{epsf}
\usepackage{CJK}
\usepackage{wasysym}
\usepackage{natbib}
\usepackage{physics}
\usepackage[T1]{fontenc}

\newcommand{\source}{ECOSMOS-LumLRD-z5}
\newcommand{\utoirac}{$u$-to-IRAC}

\received{MMDDYY}
\revised{MMDDYY}
\accepted{MMDDYY}

\submitjournal{ApJL}

\shorttitle{No Luminous LRDs}
\shortauthors{Y. Ma et al}

\graphicspath{{./}{figures/}}

\begin{document}
\begin{CJK*}{UTF8}{gbsn}
\title{No Luminous Little Red Dots: A Sharp Cutoff in Their Luminosity Function}

\correspondingauthor{Yilun Ma (yilun@princeton.edu)}
\author[0000-0002-0463-9528]{Yilun Ma (马逸伦)}
\affiliation{Department of Astrophysical Sciences, Princeton University, Princeton, NJ 08544, USA}
\email{yilun@princeton.edu}

\author[0000-0002-5612-3427]{Jenny E. Greene}
\affiliation{Department of Astrophysical Sciences, Princeton University, Princeton, NJ 08544, USA}
\email{jgreene@astro.princeton.edu}

\author[0000-0002-3216-1322]{Marta Volonteri} %last bit of fedd vs. L
\affiliation{Institut d'Astrophysique de Paris, Sorbonne Universit\'{e}, CNRS, UMR 7095, 98 bis bd Arago, F-75014 Paris, France}
\email{martav@iap.fr}

\author[0000-0003-4700-663X]{Andy D. Goulding} %done 
\affiliation{Department of Astrophysical Sciences, Princeton University, Princeton, NJ 08544, USA}
\email{goulding@astro.princeton.edu}

\author[0000-0003-4075-7393]{David J. Setton} %done 
\affiliation{Department of Astrophysical Sciences, Princeton University, Princeton, NJ 08544, USA}
\email{davidsetton@princeton.edu}

\author[0000-0002-8053-8040]{Marianna Annunziatella} %done 
\affiliation{Centro de Astrobiolog\'{\i}a (CAB), CSIC-INTA, Ctra. de Ajalvir km 4, Torrej\'on de Ardoz, E-28850, Madrid, Spain}
\email{mannunziatella@cab.inta-csic.es}

\author[0000-0003-1344-9475]{Eiichi Egami} %done 
\affiliation{Steward Observatory, University of Arizona, 933 N Cherry Avenue, Tucson, AZ 85721, USA}
\email{egami@arizona.edu}

\author[0000-0003-3310-0131]{Xiaohui Fan} %done 
\affiliation{Steward Observatory, University of Arizona, 933 N Cherry Avenue, Tucson, AZ 85721, USA}
\email{xiaohuidominicfan@gmail.com}

\author[0000-0002-5588-9156]{Vasily Kokorev} %done 
\affiliation{Department of Astronomy, The University of Texas at Austin, Austin, TX 78712, USA}
\email{} 

\author[0000-0002-2057-5376]{Ivo Labbe}
\affiliation{Centre for Astrophysics and Supercomputing, Swinburne University of Technology, Melbourne, VIC 3122, Australia}
\email{ilabbe@swin.edu.au}

\author[0000-0001-6052-4234]{Xiaojing Lin} 
\affiliation{Department of Astronomy, Tsinghua University, Beijing 100084, China}
\email{xiaojinglinastro@gmail.com}

\author[0000-0001-9002-3502]{Danilo Marchesini}
\affiliation{Department of Physics and Astronomy, Tufts University, Medford, MA 02155, USA}
\email{danilo.marchesini@tufts.edu}

\author[0000-0003-2871-127X]{Jorryt Matthee} %mostly done 
\affiliation{Institute of Science and Technology Austria (ISTA), Am Campus 1, 3400 Klosterneuburg, Austria}
\email{jorryt.matthee@ist.ac.at} 

\author[0000-0003-2804-0648]{Themiya Nanayakkara} %done
\affiliation{Centre for Astrophysics and Supercomputing, Swinburne University of Technology, Melbourne, VIC 3122, Australia} 
\email{}

\author[0000-0002-6265-2675]{Luke Robbins} %done 
\affiliation{Department of Physics and Astronomy, Tufts University, Medford, MA 02155, USA}
\email{andrew.robbins@tufts.edu} 

\author[0000-0002-1917-1200]{Anna Sajina} 
\affiliation{Department of Physics and Astronomy, Tufts University, Medford, MA 02155, USA}
\email{annie.sajina@gmail.com}

\author[0000-0002-7712-7857]{Marcin Sawicki} %done 
\affiliation{Institute for Computational Astrophysics and Department of Astronomy and Physics, Saint Mary's University, 923 Robie Street, Halifax, Nova Scotia, B3H 3C3, Canada}
\email{marcin.sawicki@smu.ca}

\begin{abstract}

One of the most surprising results of early James Webb Space Telescope (JWST) observations is the discovery of an abundance of red, compact, broad-line objects dubbed ``little red dots'' (LRDs) at $z>4$. Their spatial density ($\sim10^{-4}$--$10^{-5}\,\mathrm{cMpc^{-3}}$) is 100 times more abundant than UV-selected quasars at those redshift if one extrapolates the quasar luminosity function (QLF) down to the LRD regime. However, whether LRDs dominate black hole accretion at quasar-like luminosities ($L_\mathrm{bol}\gtrsim 10^{45-46}\,\mathrm{erg\,s^{-1}}$) remains unanswered, as probing the bright end of the LRD luminosity function requires a much larger area than those able to be surveyed by JWST. In this work, we present our search for the brightest LRDs ($K<23.7$) at $4.5<z<4.9$ using wide-area multiwavelength imaging surveys from the near-UV to the infrared bands. With over 15 square degrees of sky coverage, we only identify one single LRD candidate at $z_\mathrm{phot}\approx4.6$, which translates into a spatial density of $n(M_{5100}<-23.5)\approx10^{-8}\,\mathrm{cMpc^{-3}}$---this is nearly 10 times less abundant than the UV-selected quasars at similar optical luminosity. When combined with the LRD sample identified by JWST at the same redshift range, we find a sharp cutoff in the optical luminosity function at $\lambda L_{5100}\approx2.5\times10^{44}\,\mathrm{erg\,s^{-1}}$, while the QLF turnover occurs at $\gtrsim20$ times higher luminosity. We therefore confirm the exclusively low-luminosity nature of LRDs, ruling out that LRDs are the counter parts of quasars. Furthermore, we speculate that, if the shape of the luminosity function holds up, it points to LRDs being powered by low-mass black holes with a narrow range of Eddington-level accretion rates. 

% We argue that such a luminosity cutoff may only be reproduced if LRDs are powered by low-mass black holes ($M_\mathrm{BH}\lesssim10^7\,M_\odot$) with an Eddington ratio distribution both centered higher (likely near unity) and with less scatter than that of the UV-selected quasars. 

\end{abstract}


\section{Introduction}\label{sec:intro}

The James Webb Space Telescope (JWST) has identified a population of low-luminosity active galactic nucleus (AGN) candidates dubbed ``little red dots'' (LRDs; \citealt{Matthee2024}) across $2<z<9$. They are compact sources ($r_\mathrm{e}\lesssim100\,\mathrm{pc}$) characterized by a V-shaped spectral energy distribution (SED), where a faint blue ultraviolet (UV) continuum is followed by a steeply rising red optical continuum \citep{Kocevski2023, Akins2024, Kokorev2024lrd, Labbe2025} commonly exhibiting broad Balmer emission lines \citep[e.g.,][]{Greene2024, Juodzbalis2024rosetta, Hviding2025}. More importantly, LRDs' spectral slope change (i.e., the turnover point of the V-shaped SED) consistently occurs at rest-frame 3645\,\AA, the Balmer limit, which requires highly fine-tuned galaxy+AGN hybrid models to reproduce \citep{Setton2024inflection}. A subset of LRDs exhibit a Balmer break so strong that it is rather challenging, if not unphysical, to be explained by stellar populations alone \citep{Furtak2024qso1, Labbe2024monster, Ma2025, deGraaff2025cliff, Naidu2025bhstar, Wang2025brd}. This motivates models where the Balmer break is produced by absorption from dense $n=2$ hydrogen near the massive black hole in LRDs \citep[e.g.,][]{Inayoshi&Maiolino2025, Ji2025blackthunder}. The kinematically offset Balmer absorption that is commonly seen in LRDs may already hint the existence of such high-density gas \citep{Lin2024aspire, Matthee2024}. Beyond the UV-optical, most LRDs show no sign of strong X-ray emission from AGN corona \citep{Ananna2024, Yue2024} or infrared emission from the warm torus or galactic dust in the host galaxy \citep{Leung2024miri, Williams2024, Casey2025alma, Setton2025alma, Xiao2025noema}, suggesting that the redness may be intrinsic to the photosphere of a super-Eddington accretion flow \citep{Inayoshi2024superEdd, Kido2025, Liu2025hanpu} as opposed to dust-reddening \citep[e.g.,][]{Li2025zhengrong}. So far, no model has been able to consistently explain the full SED. 

Solving the puzzle of the SED shape of LRDs is crucial because their abundance at high redshift suggests a common phase of black hole accretion in the early universe. They are approximately 100 times more abundant than UV-selected quasars at $z>4$ \citep{Kokorev2024lrd, Kocevski2025}. However, a significant drop in LRD number density by nearly a factor of 50 occurs at $2\lesssim z\lesssim4$, and LRDs become $\sim10$ times less abundant than quasars at cosmic noon \citep{Kocevski2025, Ma2025z2lrd, Zhuang2025nexusLRD}. Although extremely rare, LRDs analogs have been identified in the local universe at $z\sim0.1$ as well \citep{Ji2025local, Lin2025local}. This suggests that the required conditions for LRDs---for instance, high gas density \citep[e.g.,][]{Inayoshi2025lognormal}---are more common, but not unique, to the high-redshift universe. 

% Yet, while the faint LRDs dominate the number density estimate, characterizing the entire luminosity function requires robust counting of the luminous ones, as one may infer the black hole properties of LRDs on a population level based on the shape of the luminosity function. Unfortunately, only a handful of luminous LRDs have been sporadically identified in different surveys, with the most luminous objects (photometrically selected candidates included) having $\lambda L_{5100} \approx 10^{45-46}\,\mathrm{erg\,s^{-1}}$ \citep{Kokorev2024lrd, Labbe2024monster, Stepney2024}. With one photometrically selected LRD candidate in their most luminous bin ($M_{1450}=-22$) within $0.18\,\mathrm{deg^2}$ of JWST blank fields, \cite{Kokorev2024lrd} tentatively claim a cutoff at $M_{1450}\approx-20.6\,\mathrm{mag}$ in the UV luminosity function (UVLF) of LRDs at $4.5<z<6.5$ beyond which the number density drops below that of quasars. Yet, due to limited survey volume of JWST and the intrinsic rarity of luminous objects, the authors cannot confidently conclude the cutoff to be physical. The bright end of the LRD luminosity function remains poorly constrained. A wide-area search is thus necessary to robustly determine its shape. Only then can we infer the properties of the central black holes and the accretion physics powering LRDs and compare them to typical blue quasars.

Yet, while the faint end of the LRD luminosity function is well characterized and dominates the number density estimation, the shape of the LRD luminosity function at the bright end remains poorly constrained. With one photometrically selected LRD candidate in their most luminous bin ($M_{1450}=-22$) within $0.18\,\mathrm{deg^2}$ of JWST blank fields, \cite{Kokorev2024lrd} tentatively claim a cutoff at $M_{1450}\approx-20.6\,\mathrm{mag}$ in the UV luminosity function (UVLF) of LRDs at $4.5<z<6.5$ beyond which the number density drops below that of quasars. However, the authors cannot confidently conclude the cutoff to be physical due to the small survey volume. A wide-area search is thus necessary to robustly determine the shape of the luminosity function on the bright end. Only then can we infer the properties of the central black holes and the accretion physics powering LRDs and compare them to typical blue quasars.

In this paper, we present our search for the most luminous LRDs using wide-area, multi-wavelength imaging surveys within 15.3 square degrees of sky to constrain the bright end of the LRD luminosity function at $4.5<z<4.9$. We introduce the photometric surveys in Section~\ref{sec:data} and the selection process in Section~\ref{sec:sample_selection}. We then present the luminosity function in Section~\ref{sec:analysis} and discuss its implication to the nature of LRDs in Section~\ref{sec:discussion}. We assume a cosmology with $H_0=70\,\mathrm{km\,s^{-1}\,Mpc^{-1}}$, $\Omega_\mathrm{m,0}=0.3$, and $\Omega_{\Lambda,0}=0.7$. All magnitudes are presented in AB magnitude \citep{Oke&Gunn1983}. 

\section{Multiwavelength Imaging Surveys}\label{sec:data}
In order to search for the most luminous LRDs, which are likely intrinsically rare, a wide-area search is required. We conduct our search within 15.3-deg$^2$ in the four Deep and UltraDeep fields (D/UD fields) of the Hyper Suprime-Cam Subaru Strategic Program (HSC-SSP; \citealt{Aihara2018hscssp}). These fields contain a wealth of multi-wavelength imaging data from the observed near-UV to the infrared wavelengths, providing sufficient wavelength coverage to select the characteristic V-shaped SED of high-redshift LRDs. Particularly, the deep HSC-$grizy$ imaging in the optical-NIR ($i_{3\sigma}\approx27.4$ and $y_{3\sigma}\approx25.8$; \citealt{Aihara2022hscdr3}) enables the detection of (and thus the selection using) the UV continuum of luminous LRDs at $m_\mathrm{UV}\lesssim26$ at $z\sim5$; see the left panel of Figure~\ref{fig:selection}).

In addition to HSC, there is also $u/u^*$-band imaging by the CFHT Large Area $U$-band Survey (CLAUDS; \citealt{Sawicki2019clauds}) in the D/UD fields. The NIR wavelengths are covered by $JHK/K_\mathrm{s}$ imaging of various ground-based surveys including the Ultra Deep Survey with the VISTA Telescope (UltraVISTA; \citealt{McCracken2012uvista}), the VISTA Deep Extragalactic Observations Survey (VIDEO; \citealt{Jarvis2013video}), the Deep eXtragalactic Survey and the Ultra-Deep Survey of the UKIRT Infrared Deep Sky Survey (UKIDSS/DXS and UKIDSS/UDS; \citealt{Lawrence2007ukidss}), the Deep UKIRT Near-infrared Steward Survey (DUNES$^2$; \citealt{Aihara2022hscdr3}, Egami et al., in prep.), and the DeepCos survey (PI: Y.-T. Lin; \citealt{Aihara2022hscdr3}). Furthermore, infrared imaging data in the 3.5 and 4.5\,$\mu$m is supplemented by the master mosaics produced by \cite{Lacy2021deepdrill} and the Spitzer Coverage of the HSC-Deep with IRAC for Z-studies 
\citep[SHIRAZ;][]{Annunziatella2023shiraz} by combining new and archival Spitzer observations. \cite{Aihara2022hscdr3} outline these multi-wavelength surveys in detail; \cite{Lacy2021deepdrill} and \cite{Annunziatella2023shiraz} also describe all the archival data invoked to produce the master IRAC mosaics that we utilize for our study. 

These multi-wavelength imaging data are then combined to construct the so-called ``\utoirac" photometric catalog. From the $u/u^*$-band to the $K/K_\mathrm{s}$-band, the catalog is defined following a modified HSC pipeline to account for non-HSC data \citep{Bosch2018hscpipe, Bosch2019lsstpipe, Desprez2023clauds+hsc}. We identify peaks in all bands simultaneously, and the detection band of a given source is assigned to be the first band in the order of $irzygJHK/K_\mathrm{s}$ that shows a $>7\sigma$ peak. We then perform forced photometry at the source position across all the ground-based bands with the same settings described in \cite{Aihara2022hscdr3}. For the Spitzer/IRAC data, photometry is carried out using a de-blending algorithm. Source segmentation maps are constructed using the $K/K_\mathrm{s}$-band, $J$-band, and $z$-band imaging data---redder bands have higher priority---and used as priors to build models with IRAC resolutions. The nearby and/or blended sources are then subtracted and photometry is carried out using a $D=3\,''$ aperture. This de-blending procedure is described in detail in \cite{Marchesini2009smf}. We subsequently correct the aperture fluxes to total fluxes using a median growth curve based on point sources identified in each field's mosaic. 

Lastly, for this work specifically, we remove objects that only have peaks detected in less than three bands and objects that are blended with bright stars. These sources are not involved in estimating the total survey area.

\section{Sample Selection}\label{sec:sample_selection}

\subsection{LRD Selection}\label{sec:selection}

\begin{figure*}[ht!]
    \centering
    \includegraphics[width=\textwidth]{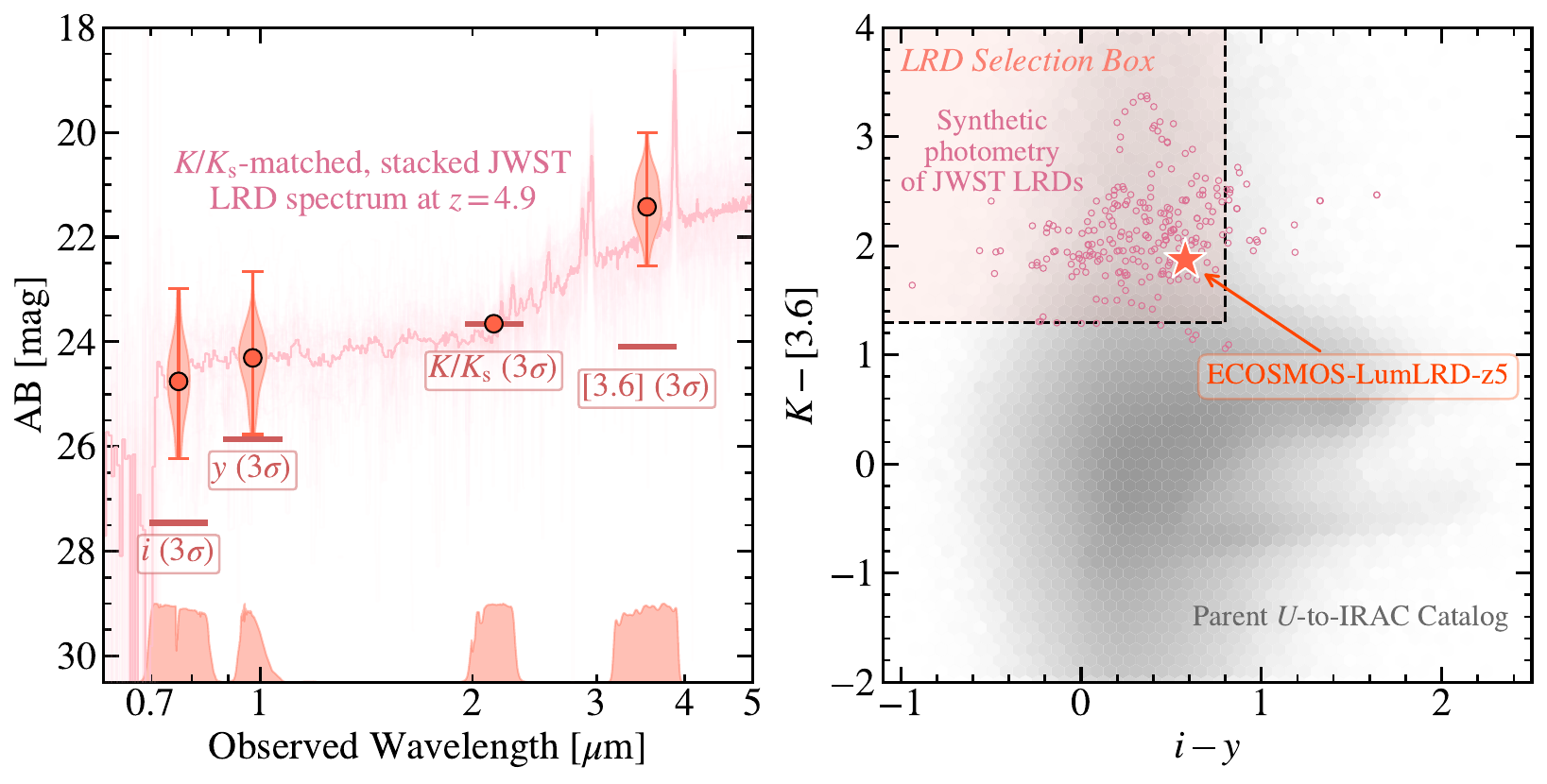}
    \caption{\textbf{Left:} The spectra of JWST LRDs from \cite{Setton2024inflection} and \cite{Hviding2025} are shifted to $z=4.9$, matched to the $K/K_\mathrm{s}$-band $3\sigma$ limit (23.7\,mag), and stacked together. Synthetic photometry of the stack (circles) and of each individual LRD in the stack (violins) both demonstrate that, after applying the $K/K_\mathrm{s}$ cut, the photometry of all selection bands would lie above their respective survey depth limits (horizontal lines), suggesting our search for the luminous LRDs is plausibly volume-complete. \textbf{Right:} The synthetic photometric colors of 38 spectroscopically selected LRDs from \cite{Setton2024inflection} and \cite{Hviding2025} are shown as pink circles in the LRD selection box. The colors are computed by shifting the 38 LRD spectra to $4.5<z<4.9$ with a $\Delta z=0.1$ increment. The one LRD candidate (\source) that we identify this work is shown as the star. The full parent \utoirac\ catalog is shown in the background. }
    \label{fig:selection}
\end{figure*}

The selections for LRDs, either using photometric colors \citep[e.g.,][]{Akins2024, Labbe2025, Ma2025z2lrd, Hviding2025} or using photometric slopes \citep{Kocevski2025}, are all designed to target their characteristic V-shaped UV-optical SEDs. We invoke the following criteria for our selection of LRDs:
\begin{enumerate}
    \item the object must be detected with a signal-to-noise ratio (SNR) greater than 3 from $r$ to [4.5];
    \item the object must be a drop-out ($\mathrm{SNR}<3$) in $g$-band;
    \item $K/K_\mathrm{s}<23.7$;
    \item $i-y<0.8$;
    \item $K/K_\mathrm{s}-[3.6]>1.3$;
    \item the object must be consistent with a point source (represented by \texttt{extendedness=0} as defined in the HSC pipeline; see \citealt{Bosch2018hscpipe}) in $i$-, $J$-, $H$-, $K/K_\mathrm{s}$-bands.
\end{enumerate}

The drop-out criterion ensures that we target the redshift range of interest. The $K/K_\mathrm{s}$-band magnitude cut is set to be the shallowest $K/K_\mathrm{s}$-band $3\sigma$ depth for blind point source detection among the HSC-D/UD fields, targeting the brightest objects only. As shown in the left panel of Figure~\ref{fig:selection}, the magnitude cut also makes the brightness of the other three selection bands lie above their respective depths ($i_{3\sigma}=27.4$, $y_{3\sigma}=25.8$, and [3.6]$_{3\sigma}=24.1$)\footnote{Like the $K$-band, these depths are also the shallowest $3\sigma$ blind point source detection limits in each band over the full survey volume.}. The $i-y$ color targets the blue UV continuum of the LRDs, while the $K/K_\mathrm{s}-[3.6]$ color targets the red optical continuum. In the right panel of Figure~\ref{fig:selection}, we show that the synthetic colors of many spectroscopically selected JWST LRDs also reside in the selection box. We give a detailed discussion on the selection completeness in Section~\ref{sec:completeness}.

Leveraging the high spatial resolution of space-based observations, LRD studies with JWST often quantify compactness using photometry or morphology within the innermost 1--2\,kpc of the sources in the rest-frame optical bands \citep[e.g.,][]{Hviding2025, Kocevski2025, Labbe2025}. It is worth noting that extended morphology in the rest-frame UV has been uncovered in a fraction of JWST LRDs, suggesting potential host or gas nebular emission \citep[e.g.,][]{Rinaldi2024, Chen2025host, Torralba2025monster_muse}. Our compactness requirement is not as stringent as that imposed by JWST studies due to the nature of seeing-limited observations ($0.^{\prime\prime}66$ in $i$-band corresponds to $\sim4$\,kpc at $z=4.7$; \citealt{Aihara2022hscdr3}) from the ground---we discuss this in the case-by-case situation in the following paragraph. 

Using the aforementioned six criteria, we only identify two potential LRD candidates from the \utoirac\ parent sample within the entire survey area. A breakdown of number of objects included by each selection criterion can be found in Appendix~\ref{app:selection}. We now attempt to visually inspect higher-resolution imaging data of the two candidates. One of the two candidates (RA$\,=\,$09:59:39.12, Dec$\,=\,$+02:06:01.48) is within the imaging footprints of the COSMOS-Web survey \citep{Casey2023cosmosweb} and Public Release Imaging for Extragalactic Research survey (PRIMER; GO \#1837; PI: J. Dunlop). The candidate clearly displays an extended disk in the NIRCam/F444W image. It is true that recent work by \cite{Rinaldi2025} identifies an extended galaxy with an LRD-like nucleus. Although such objects may hint at the evolutionary path of LRDs, they do not resemble typical LRDs selected by JWST \citep[e.g.,][]{Akins2024, Kokorev2024lrd, Kocevski2025} that we use to combine our sample with (see Section~\ref{sec:olf}). Thus, we exclude this JWST-extended source, and only one single object remains.  

Lastly, brown dwarfs, which also appears as point sources at JWST-resolution, could serve as contamination when selecting LRDs at $z\gtrsim4$ using photometric colors of JWST \citep{Burgasser2024, Greene2024, Labbe2025}, but this is not the case for us. It is because the $i-y$ colors of brown dwarfs are much redder than the NIRCam colors used to select the blue rest-frame UV of LRDs (e.g., F115$-$F200W), as the former broadly traces the Wien tail and the red wing of the resonant K\,{\sc i} absorption of brown dwarf SEDs. Thus, we conclude that the remaining LRD candidates is extragalactic in nature. This luminous LRD candidate, \source, is the only source we identify among nearly two million sources over the 15.3-square-degree survey field. 

\subsection{\source}
\begin{figure}[ht]
    \centering
    \includegraphics[width=\columnwidth]{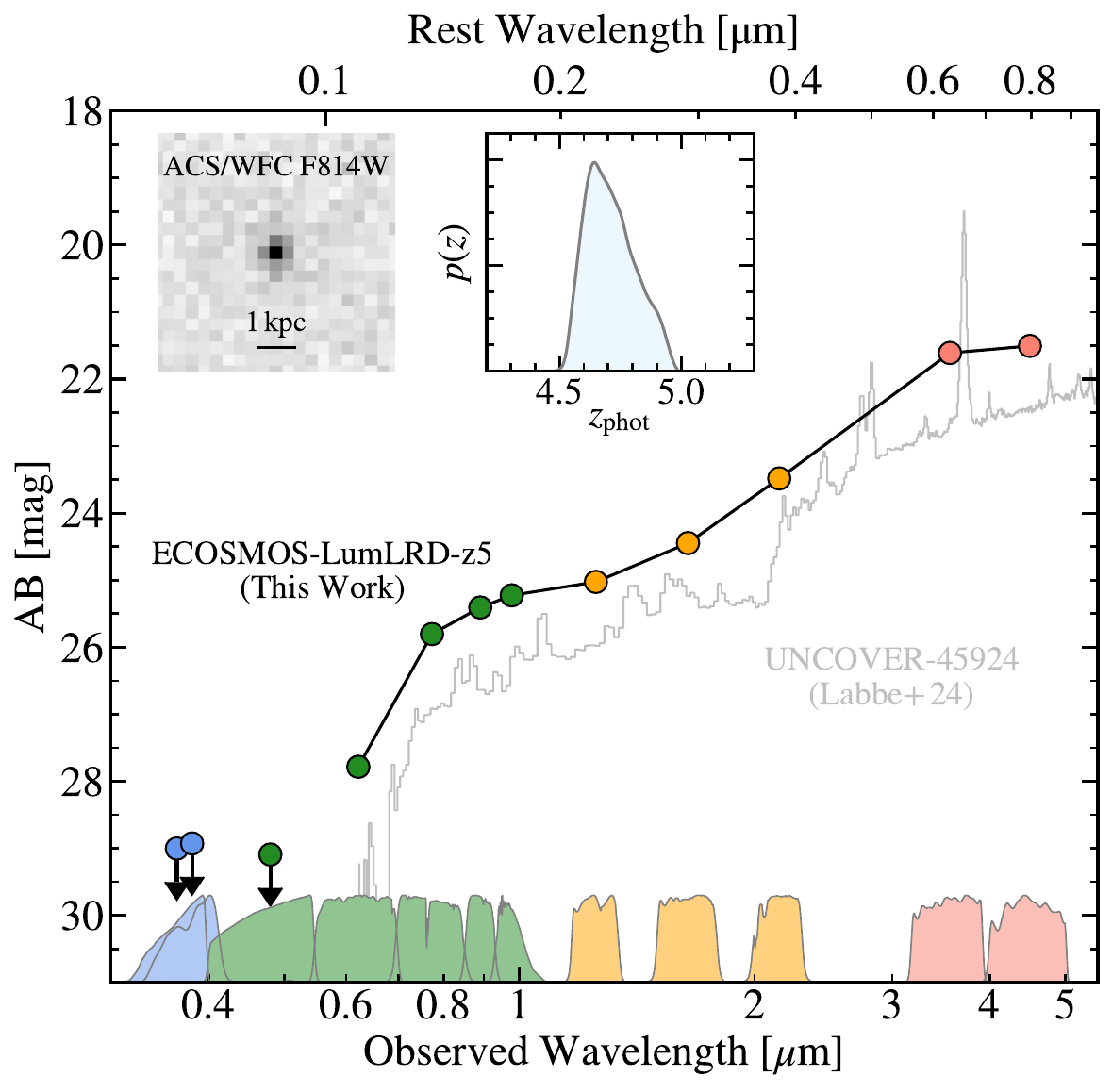}
    \caption{The photometric SED of \source\ is shown against the NIRSpec/PRISM spectrum of UNCOVER-45924 \citep{Labbe2024monster}, the most luminous JWST LRD to date. The HST/ACS/WFC F814W image cutout and the $p(z)$ distribution output by \texttt{EAZY} are shown in the insets.}
    \label{fig:sed}
\end{figure}

\begin{deluxetable}{lll}[ht]
\tablecaption{Source Properties of \source 
\label{tab:source}}
\tablewidth{0pt}
\tablehead{
\colhead{Quantity} & \colhead{Unit} & \colhead{Measurements}
}
\startdata
RA & deg & $149.8664$ \\
Dec & deg & $+1.7497$ \\
$z_\mathrm{phot}$ & --- & $4.65_{-0.04}^{+0.18}$ \\
\hline
CFHT/MegaCam--$u$ & mag & $>29.00$ \\
CFHT/MegaCam--$u^*$ & mag & $>28.93$ \\
Subaru/HSC--$g$ & mag & $>29.09$ \\
Subaru/HSC--$r$ & mag & $27.78\pm0.15$ \\
Subaru/HSC--$i$ & mag & $25.80\pm0.03$ \\
Subaru/HSC--$z$ & mag & $25.41\pm0.03$ \\
Subaru/HSC--$y$ & mag & $25.23\pm0.05$ \\
VISTA/VIRCAM--$J$ & mag & $25.03\pm0.02$ \\
VISTA/VIRCAM--$H$ & mag & $25.45\pm0.02$ \\
VISTA/VIRCAM--$K_\mathrm{s}$ & mag & $23.48\pm0.01$ \\
Spitzer/IRAC\,[3.6] & mag & $21.61\pm0.02$ \\
Spitzer/IRAC\,[4.5] & mag & $21.50\pm0.03$ \\
\enddata
\end{deluxetable}

\source\ is the only LRD candidate brighter than $K/K_\mathrm{s}=23.7$ that we identified in our search. Its position in the color-color space is shown in the right panel of Figure~\ref{fig:selection}.  We show its multiwavelength SED in Figure~\ref{fig:sed} and Table~\ref{tab:source}. We estimate the photometric redshift of \source\ with \texttt{EAZY} \citep{Brammer2008eazy} and obtained $z_\mathrm{phot}=4.65_\mathrm{-0.04}^\mathrm{+0.18}$. The $p(z_\mathrm{phot})$ distribution is shown in Figure~\ref{fig:sed}, and we point the readers to Appendix~\ref{app:eazy} for the details of the \texttt{EAZY} configuration we use. With some handle on the source redshift, we note that \source\ is nearly $\sim1$\,mag more luminous than UNCOVER-45924, the most luminous LRD discovered and spectroscopically confirmed by JWST to date \citep{Labbe2024monster}. 

\source\ has no JWST imaging available but is within the Hubble Space Telescope (HST) footprints of the Cosmic Evolution Survey (COSMOS; \citealt{Koekemoer2007cosmoshst, Scoville2007cosmos, Massey2010cosmos_hst_reduction}). The ACS/F814W imaging cutout of \source\ is also shown in Figure~\ref{fig:sed}. We provide the details of the morphological analysis in Appendix~\ref{app:source_morphology} and conclude that \source\ is consistent with a point source in the rest-frame UV. We thus estimate a size upper limit of $r<310\,\mathrm{pc}$ for \source, similarly compact to JWST-selected LRDs \citep[e.g.,][]{Akins2024}.

\subsection{Selection Completeness}\label{sec:completeness}

We examine the completeness of our selection with a set of spectroscopically confirmed LRDs at $3.9<z<7.0$ from \cite{Setton2024inflection} and \cite{Hviding2025} with spectra publicly available on the Dawn JWST Archive\footnote{\href{https://dawn-cph.github.io/dja/index.html}{https://dawn-cph.github.io/dja/index.html}}. Five RUBIES objects (EGS-53254, EGS-37124, UDS-29813, UDS-149298, and UDS-807469) are within the redshift range but removed because of their incomplete wavelength coverage by NIRSpec/PRISM. We exclude another three RUBIES objects (EGS-42232, UDS-36171, UDS-830237) as we determine their rest-frame UV to be too faint for accurate synthetic photometry after our visual inspection. The remaining 38 objects are referred to as ``templates" in the following discussion. 

We first shift the template spectra to $z=4.9$, the upper bound of our redshift range of interest. We compute the synthetic photometry of each object in $i$, $y$, $K$, and [3.6] bands. After matching all templates to $K=23.7\,\mathrm{mag}$ among the entire 15.3-deg$^2$ search area, we stack all templates and measure the four-band synthetic photometry on the stack. In the left panel of Figure~\ref{fig:selection}, we show that once the $K$-band cut is made, the brightness of the stack and of each individual template in the other three bands are all above their respective depths. This heuristically suggests that our search of the brightest LRDs is plausibly volume-complete for the typical objects. 

To investigate the completeness in details, we create mock observations using the template spectra. For each template, we randomly draw redshift from $z\in[4.5, 4.9]$ and monochromatic luminosity at rest 5100\,\AA\ from $M_{5100}\in[-23.5, -24.5]$ with 500 realizations---the latter is the luminosity bin we are interested in (see Section~\ref{sec:olf}). We shift the template spectra to the observed frame according to the redshift and luminosity and compute synthetic photometry of the mock objects from $i$-band to and IRAC/[3.6]. Finally, the synthetic photometry $m_\mathrm{syn}$ in each band is perturbed according to an uncertainty randomly drawn from the error distribution derived from sources within $m_\mathrm{syn}\pm0.5$ in the parent \utoirac\ catalogs. We repeat the above procedure on all 38 LRD template spectra and feed all ($38\times500=19,000$) mock LRD photometries into the selection pipeline (SNR cut, magnitude cut, and V-shape), recovering only 11,467 mock sources. Thus, we estimate the selection completeness to be $C_\mathrm{sel}=0.60$ for our ground-based search. This is slightly higher than the 50\% completeness estimated by \cite{Hviding2025} for the JWST color-color selection of \cite{Kokorev2024lrd}. This difference is likely due to the fact that our search is not as limited by the UV faintness as the JWST selection does because the $K$-band, which traces the rest-frame optical, is the shallowest. 

\section{Number Density Estimation}\label{sec:analysis}

\subsection{Survey Volume}\label{sec:volume}
First of all, we comment on the redshift range, $4.5<z<4.9$, that we restrict ourselves to when estimating the number density. In fact, our $i-y$ vs. $K-[3.6]$ criteria is not entirely insensitive to LRDs at $3.8<z<4.5$ and $4.9<z<5.2$. Our choice of $4.5<z<4.9$ is to enable a fair comparison with color-selected JWST samples (which adopt a redshift lower bound of $z=4.5$; \citealt{Kokorev2024lrd}) and to ensure that the $i$-band flux is unaffected by the IGM absorption (including the damped wing of Ly$\alpha$ absorption).

As a result, the volume of interest should in principle be larger than the comoving volume between $z=4.5$ and $z=4.9$ within the survey area. Yet, we think that estimating the volume using the full $3.8<z<5.2$ range would only result in an order-unity effect to our subsequent number density estimation, as the comoving volume only increases by a factor of $\sim3.5$. Moreover, since  the number density of LRDs rapidly drops by a factor of $\sim3$--5 at $3.8<z<4.5$ \citep{Inayoshi2025lognormal, Ma2025z2lrd, Tanaka2025z10lrd} and our selection completeness drops to $\sim20\%$ at $4.9<z<5.2$, any unidentified sources at those higher and lower redshifts are all consistent with $\sim$Poisson(1) once the order-unity volume increase is accounted for. Lastly, photometric selections are subject to contamination, which could lower the true number density, achieving similar order-unity effect as an increased survey volume. 

% As a result, the volume of interest should in principle be larger than $\sim V_\mathrm{comoving}(4.5<z<4.9)$. However, we consider this an order-unity effect to our number density estimation. Firstly, our measured number density may be regarded as an upper limit. Since we do not exclude any objects from our sample based on photometric redshift measurements (not that we are able to given a single identified source), if we use the full redshift range, our estimated number density would decrease only by a factor of $\sim3.5$ due to the volume increase, fully consistent with the $1\sigma$ uncertainty of the measured value. In addition, the number density could also decrease if spectroscopic follow-up of \source\ rules out its LRD nature. Secondly, we only expect $\sim1--2$ more objects if we were to include any potential source below $z=4.5$ or above $z>4.9$ given the 50\% incompleteness of color-color selection \citep{Hviding2025} and the redshift evolution of LRDs \citep{Inayoshi2025lognormal, Ma2025z2lrd, Tanaka2025z10lrd}. In the meanwhile, the increased volume would also keep the then-measured number density unchanged in bulk part. 
%Moreover, given the redshift evolution of LRDs \citep{Inayoshi2025lognormal, Ma2025z2lrd, Tanaka2025z10lrd}, we may expect to find 1--2 more LRD candidates the full redshift range, but the volume would also increase by the same factor, keeping the measured number density unchanged in bulk part. Lastly, we do not remove objects (not that we are able to given one single identified candidate). 

\subsection{Optical Luminosity Function}\label{sec:olf}

\begin{deluxetable}{ccc}[ht!]
\tablecaption{The optical luminosity function of LRDs at $4.5<z<4.9$ \label{tab:lf}}
\tablewidth{0pt}
\tablehead{
\colhead{Sample} & \colhead{$M_{5100}$} & \colhead{$\log\Phi$} \\ 
\colhead{} & \colhead{[mag]} & \colhead{[$\mathrm{cMpc^{-3}mag^{-1}}$]}
}
\startdata
This Work & $-25$ & $<-7.53$ \\
{} & $-24$ & $-7.57_{-0.76}^{+0.52}$ \\\hline
{} & $-23$ & $-5.85_{-0.76}^{+0.30}$ \\
{} & $-22$ & $-5.01_{-0.24}^{+0.15}$ \\
\cite{Kokorev2024lrd} & $-21$ & $-4.49_{-0.11}^{+0.14}$ \\
{} & $-20$ & $-4.31_{-0.19}^{+0.20}$ \\
{} & $-19$ & $-4.72_{-0.23}^{+0.19}$ \\
\enddata
\tablecomments{The magnitude bins all have a width of $\pm0.5\,\mathrm{mag}$, and the upper limit for the highest luminosity bin is given as the $1\sigma$ value.}
\end{deluxetable}

\begin{deluxetable}{ccc}[ht!]
\tablecaption{Best-Fit Model Parameters for the optical luminosity function of LRDs at $4.5<z<4.9$ \label{tab:schechter}}
\tablewidth{0pt}
\tablehead{
\colhead{Parameter} & \colhead{Units} & \colhead{Value}}
\startdata
{} & Schechter Function & {} \\\hline 
$\Phi^*$ & $\mathrm{10^{-5}\,cMpc^{-3}}$ & $3.95_{-2.05}^{+5.45}$ \\
$M^*$ & mag & $-21.9_{-0.3}^{+0.9}$ \\
$\alpha$ & --- & $-1.30_{-0.42}^{+0.87}$ \\\hline 
{} & Double Power Law & {} \\\hline 
$\Phi^*$ & $\mathrm{10^{-5}\,cMpc^{-3}}$ & $0.83_{-0.63}^{+4.33}$ \\
$M^*$ & mag & $-22.8_{-1.0}^{+1.2}$ \\
$\alpha$ & --- & $-1.74_{-0.37}^{+0.86}$ \\
$\beta$ & --- & $-6.13_{-19.93}^{+1.80}$
\enddata
\tablecomments{$M_{5100}^*=-21.9$\,mag and $-22.8$\,mag can be converted into $\lambda L_{5100}=1.5\times10^{44}$\,erg\,s$^{-1}$ and $3.4\times10^{44}$\,erg\,s$^{-1}$, respectively.}
\end{deluxetable}

\begin{figure}[ht]
    \centering
    \includegraphics[width=\columnwidth]{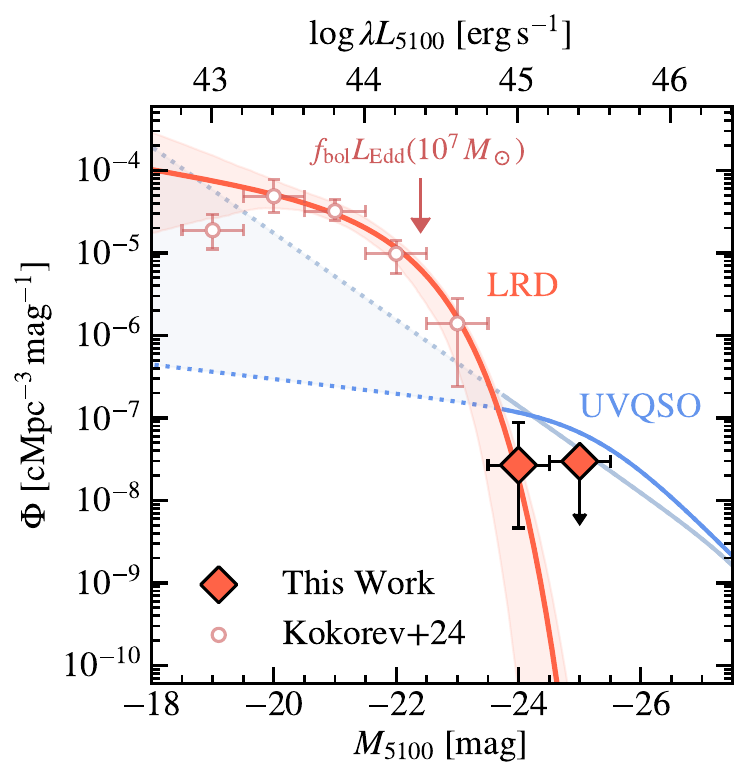}
    \caption{\textbf{Left:} The optical luminosity function of LRDs at $4.5<z<4.9$ is shown in red. The solid diamonds are the measurements done in this work.  The upper limit is the $1\sigma$ value. The empty circles are measurements from LRD candidates identified by \cite{Kokorev2024lrd} at $4.5<z<4.9$ with JWST. The best-fit Schechter model is shown in the red with the 68\% confidence interval shaded. The light and dark blue curves are the LFs for UV-selected quasars at $4.7<z<5.5$ measured by \cite{Kulkarni2019QLF} and at $z\sim5$ by \cite{Niida2020QLF}. The dashed lines are the extrapolated quasar LFs down to the faint ends. The arrow marks the optical luminosity of a $10^7\,M_\odot$-black hole with $L_\mathrm{bol}/L_\mathrm{Edd}=1$, assuming $L_\mathrm{bol}/\lambda L_{5100}=5.4$ for the LRDs (J. E. Greene et al., submitted). In contrast, the bolometric correction for the quasars is $L_\mathrm{bol}/\lambda L_{5100}=10.3$ \citep{Richards2006}. }
    \label{fig:LF}
\end{figure}

Following \cite{Ma2025z2lrd}, we also choose to compute the optical luminosity function, parametrized by the absolute magnitude at rest-frame 5100\,\AA\ ($M_{5100}$), instead of the UV luminosity function (UVLF). While the origin of LRDs' UV continuum remains unclear as only a fraction of objects shows extended morphology \citep{Rinaldi2024, Chen2025nebular, Chen2025host, Torralba2025monster_muse}, the origin of the optical continuum appears to be converging towards a consensus. In models where a gas cocoon surrounds the accreting black hole and produces the red optical continuum \citep[e.g.,][]{Liu2025hanpu, Naidu2025bhstar, Rusakov2025}, the optical luminosity likely provides a more sensible tracer of the central engine's accretion power (J. E. Greene et al., submitted) with minimal host emission like the UV. This makes the optical luminosity function particularly advantageous over the UVLF. 
%Lastly, this practice reduces the selection effects that the UV luminosity may be strongly dependent upon the red optical color of LRDs \citep{Ma2025z2lrd}. For example, at a given optical luminosity, the redder an object is in the optical, the more likely it to be fainter in the UV. 

Given the identification of a single LRD candidate, we compute the number density of LRDs within a luminosity bin $\Delta M$ with 
\begin{equation}
    \Phi(M)\Delta M = \frac{1}{V_\mathrm{max}\times C_\mathrm{sel}}\,\,,
\end{equation}
where $C_\mathrm{sel}$ is the selection completeness discussed in Section~\ref{sec:completeness}, and $V_\mathrm{max}$ is the maximum comoving volume that a source with a given luminosity can be detected in a magnitude-limited survey \citep{Schmidt1968}. The luminosity function is computed with 1000 Monte Carlo realizations. Within each realization, the photometry is drawn based on its uncertainty and redshift from the $P(z)$ distribution to compute $M_\mathrm{5100}$ by interpolating the photometric SED and to estimate $V_\mathrm{max}$. Then, we combine the Monte Carlo uncertainties of the number density with the Poisson uncertainties \citep{Gehrels1986}. 

In our specific case, we take the smallest of the $V_\mathrm{max}$ estimator in all four bands used for color selection as the effective volume---this is similar to the approach by \cite{Kokorev2024lrd}. We find that $V_\mathrm{max}$ is just equal to the comoving volume between $z=4.5$ and $z=4.9$ within the 15.3-deg$^2$ search area---this is not surprising as our search for luminous LRDs is likely volume-complete within this redshift range. We therefore obtain the number density within $-24.5<M_{5100}<-23.5$ to be $10^{-7.57}\,\mathrm{cMpc^{-3}}$. Again, as mentioned in Section~\ref{sec:volume}, using the full $3.8<z<5.2$ range over which our selection is sensitive would only increase the volume by a factor of $\sim3$, further suppressing number density at the bright end of the LF. 
%As a caveat, we note that since \source\ requires spectroscopic follow-up to confirm its true LRD nature and since our selection may also select some LRDs at $3.8<z<4.5$ and $4.9<z<5.2$ as we mention in Section~\ref{sec:selection}, this measured number density may be regarded as an upper limit. 
We also add one more luminosity bin at $-25.5<M_{5100}<-24.5$ and compute the $1\sigma$ and $3\sigma$ Poisson upper limits of $<10^{-7.53}\,\mathrm{cMpc^{-3}\,mag^{-1}}$ and $<10^{-7.07}\,\mathrm{cMpc^{-3}\,mag^{-1}}$, respectively. 

Although our ground-based search grants a large area for the rare and luminous LRDs, it is not deep enough for the fainter ones ($K/K_\mathrm{s}\gtrsim24$). Therefore, we invoke the LRD sample in blank deep JWST/NIRCam fields presented in \cite{Kokorev2024lrd}. This particular sample is chosen because the authors employ photometric colors very similar to ours for sample selection, which ensures consistency when bridging our sample with the JWST one. Using the photometric-slope-selected sample of \cite{Kocevski2025} would likely not significantly affect our conclusion given the similarity in both works' number density estimations of less luminous LRDs. Utilizing the photo-$z$ from \cite{Kokorev2024lrd}, we calculate the $M_{5100}$ of the JWST LRDs whose redshift is consistent with the redshift range of interest within $1\sigma$ in a similar fashion. This fills in the faint end of the optical LF of LRDs at $4.5<z<4.9$. We present the optical luminosity function constructed from both samples in Table~\ref{tab:lf}. 

Similar to \cite{Kokorev2024lrd}, we model the luminosity function in the form of a Schechter function \citep{Schechter1976}, 
\begin{align}
    \Phi(M)\dd M = (0.4\ln10)\times\Phi^*\times10^{0.4(\alpha+1)(M^*-M)} \notag&\\\times\exp\left[-10^{0.4(M^*-M)}\right]\dd M&\,\,,
\end{align}
where $M$ is the absolute magnitude, $M^*$ is the characteristic absolute magnitude determining the position of the exponential drop-off at the bright end, $\alpha$ controls the power-law slope at the faint end, and $\Phi^*$ is a normalization. Again, the uncertainties are propagated in a Monte Carlo fashion with 1000 realizations. We show the best-fit model and its parameters in Figure~\ref{fig:LF} and Table~\ref{tab:schechter}, respectively. 

To compare the LRDs with other accreting black holes, we convert the UVLFs of UV-selected quasars (UVQSOs) at comparable redshift into optical LFs as well. The optical LFs of the UVQSOs are approximated by estimating $M_{5100}-M_{1450}=-0.26\,\mathrm{mag}$ from an empirical $z=4.7$ quasar template derived by \cite{Temple2021} and shifting the UVLF by this constant offset. In Figure~\ref{fig:LF}, it is clear that LRDs quickly disappear above $M_{5100}\lesssim-22.5$, while the UVQSOs display a luminous tail in their luminosity functions extending to at least two or three magnitudes brighter than the luminosity cutoff of the LRDs. At $M_{5100}\lesssim-24$, the UVQSOs are more abundant than LRDs by at least 1\,dex. Although LRDs seem to dominate the black hole number density at $z\sim4.5$, UVQSOs still dominate the most luminous episodes of black hole accretion. Furthermore, the stark difference between the LF shapes of the LRDs and the UVQSOs may suggest that LRDs do not represent the simple obscured counterparts of luminous quasars. 

Finally, we make sure the steep cutoff in the LRD luminosity function that we claim is not due to the exponential term in the Schechter function. As a sanity check, we also model the measured number densities with a double power law \citep{Boyle1988}, 
\begin{equation}
    \Phi(M)\dd M = \frac{\Phi^*\dd M}{10^{0.4(\alpha+1)(M-M^*)}+10^{0.4(\beta+1)(M-M^*)}}\,\,,
\end{equation}
where $\alpha$ and $\beta$ are the faint- and bright-end slopes, respectively---this parameterization is commonly used to describe quasar luminosity functions (QLFs). The best-fit parameters of the double power law are also listed in Table~\ref{tab:schechter}. We measure a bright-end slope of $\beta=-6.13_{-19.93}^{+1.80}$ for the LRDs. In comparison, the bright and faint end slopes for the UVQSOs are $\beta_\mathrm{QSO}\sim-4.5$ and $\alpha_\mathrm{QSO}\gtrsim-2$ \citep{Kulkarni2019QLF, Niida2020QLF}, respectively. Since both QLF slopes are shallower than those of the LRDs that we measure, we therefore conclude the steep luminosity cutoff for the LRDs is likely physical. 

\section{What Types of Black Holes are They?}\label{sec:discussion}

\begin{figure*}[ht]
    \centering
    \includegraphics[width=\textwidth]{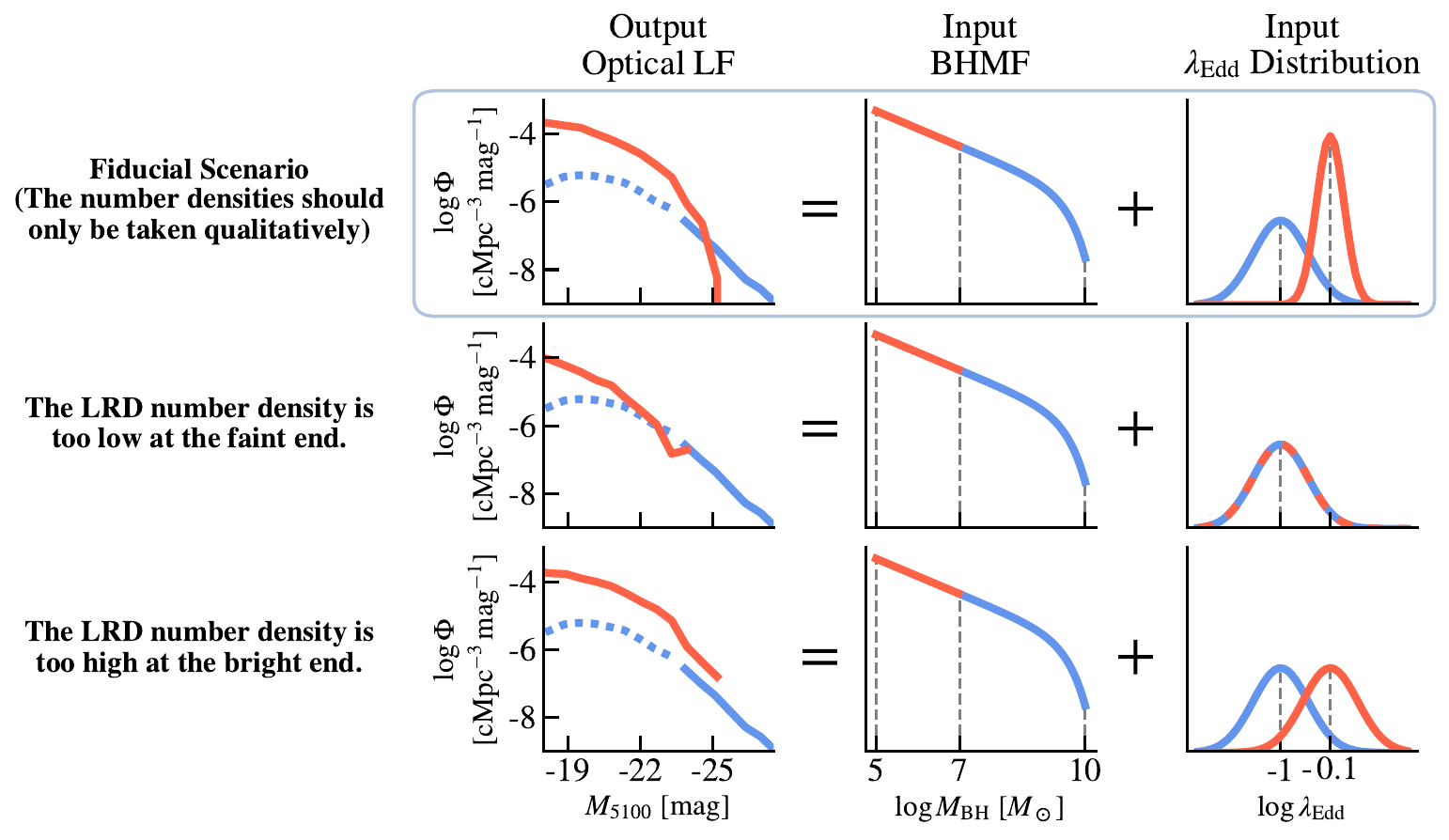}
    \caption{A schematic cartoon showing how different combinations of black hole mass functions and Eddington
    ratio distributions of the LRDs (red) and quasars (blue) affects the shape of the luminosity functions. The first row is our fiducial model, where LRDs represent a population of low-mass black holes with high Eddington ratio and small scatter in the Eddington ratio distribution. For the quasars, the quasar luminosity function is extrapolated below an observed limit and marked with the dashed curve. This figure is entirely demonstrative and should mostly be taken qualitatively.}
    \label{fig:sims}
\end{figure*}

Despite using a UVLF, \citet{Kokorev2024lrd} already speculate about the lack of luminous LRDs in their search across four JWST blank fields. Nonetheless, they caution that this finding may be subject to the limited survey area---0.18 square degrees, compared to the 15.3 square degrees covered in our search---and may not reflect a physical deficit. With our significantly larger area, we robustly establish that the number density of LRDs declines sharply beyond $M_{5100} \lesssim -22.5$\,mag. As mentioned in previous sections, selection incompleteness or expanding the redshift range used to compute the volume could not affect the conclusion on order-of-magnitude level. Rather than using a UVLF, we focus on the optical luminosity as a more physically motivated tracer of intrinsic properties of the central engine. Given this, we now turn to the question of why LRDs exhibit such a steep luminosity cutoff at relatively modest luminosities compared to UV-selected quasars, which are similarly powered by accretion onto massive black holes.

In the simplest terms, the luminosity of an accreting black hole scales with the product of the Eddington ratio ($\lambda_\mathrm{Edd}\equiv L/L_\mathrm{Edd}$) and the black hole mass, i.e., $L\propto \lambda_\mathrm{Edd}M_\mathrm{BH}$. We therefore center our discussion---and the accompanying heuristic illustrations (see Figure~\ref{fig:sims})---on these two physical quantities.

Seemingly, the most straightforward hypothesis for such a steep luminosity cutoff is that LRDs are uniquely a low-$M_\mathrm{BH}$ phenomenon when compared to the quasars. This is particularly true given that LRDs may have large $\lambda_\mathrm{Edd}$ values. Super-Eddington accretion has been commonly invoked to explain the X-ray weakness \citep[e.g.,][]{Lambrides2024, Pacucci&Narayan2024} and the abundance of LRDs at $z\gtrsim4$ \citep[e.g.,][]{Greene2024, Trinca2024}. Using bolometric corrections measured from LRDs, $f_\mathrm{bol}=\lambda L_{5100}/L_\mathrm{bol}=0.18$ (J. E. Greene et al., submitted), we estimate the bright-end cutoff of the LRDs roughly matches the optical luminosity of a $\sim10^7\,M_\odot$ black hole with $L_\mathrm{bol}=L_\mathrm{Edd}$. This is $\sim50$--100 times less massive than the typical black hole masses inferred from UV-selected quasars at comparable redshifts \citep{He2024hscqsoz4}. Such a faint luminosity cutoff in LRDs relative to quasars may thus be interpreted as a cutoff in the underlying black hole mass as well. 

Additionally, the bright end of the LRD luminosity function is steeper than that of the quasars. This could result from a narrower Eddington ratio distribution for LRDs (e.g., well clustered at $L_\mathrm{bol}/L_\mathrm{Edd}\approx1$) relative to quasars ($0.1\lesssim L_\mathrm{bol}/L_\mathrm{Edd}\lesssim1$; \citealt{He2024hscqsoz4}, \citealt{Li2024fedd}). Such a ``narrower" Eddington ratio distribution would make low-mass accretors less likely to be scattered into the high-luminosity tail of the whole population. In fact, less scatter in the Eddington ratio distribution may be a natural outcome for super-Eddington accretion. Theoretical works suggest $L_\mathrm{bol}\sim \mathcal{O}(1-10)\times L_\mathrm{Edd}$ even when the accretion rate exceeds the Eddington value, potentially due to significant photon trapping at sufficiently high accretion rate \citep{Abramowicz1988slimdisk, Watarai2000mdotL, Jiang2014, Sadowski2015mdotL, Inayoshi2020araa}. Moreover, if we assume $L_\mathrm{bol}\sim L_\mathrm{Edd}$ is required to power an LRD and that the accretion is fuel-limited, then an upper bound on $M_\mathrm{BH}$ ensues. Put another way, a high Eddington ratio is more likely to occur when black hole masses are low.

Motivated by those arguments, we heuristically test the effect of an $M_\mathrm{BH}$ cutoff and a narrow $\lambda_\mathrm{Edd}$ distribution centered close to unity on the shape of the luminosity function of the LRDs using the semi-empirical framework outlined by \cite{Volonteri2017}. Briefly, we assign a black hole with $M_\mathrm{BH}$ in every galaxy according to the $M_\mathrm{BH}$--$M_\star$ relation \citep{Greene2020} with scatter $\sigma_{\log M_\mathrm{BH}}=0.5$ (for both LRDs and the quasars) after sampling the galaxies from a synthetic stellar mass function at $z=5$ (see \citealt{Dayal&Giri2024} for details of the stellar mass function). Assuming a duty cycle of 0.1 and an obscured fraction of 90\%, we sample from a log-normal Eddington ratio distribution centered at $\langle\log\lambda_\mathrm{Edd}\rangle$ with a dispersion $\sigma_{\log\lambda_\mathrm{Edd}}$ to assign luminosity to each black hole, which can then be used to estimate an LF. In this heuristic exercise, we define LRDs as systems with $M_\mathrm{BH}<10^7\,M_\odot$ (motivated by our measured cutoff luminosity; see Figure~\ref{fig:LF}) with $\langle\log\lambda_\mathrm{Edd}\rangle=\log0.8$ and $\sigma_{\log\lambda_\mathrm{Edd}}=0.25$, while quasars as those with $M_\mathrm{BH}>10^7\,M_\odot$ with $\langle\log\lambda_\mathrm{Edd}\rangle=\log0.1$, $\sigma_{\log\lambda_\mathrm{Edd}}=0.50$, and $M_{5100}<-23.5\,$mag (similar to the the fainter limit in observed QLFs (see Figure~\ref{fig:LF}). We refrain from a full exploration of the parameter space in this work, as the duty cycle, obscured fraction, black hole occupation fraction, the black hole mass of LRDs, and scaling relations at $z=5$ are not constrained in reality. Thus, we stress that our tests here are purely heuristic and should only taken qualitatively. We show the result of this heuristic test in the upper row of Figure~\ref{fig:sims}. Indeed, qualitatively, we are able to reproduce the steep drop of LRD LF at a fainter luminosity than that of quasars with this toy model. 

We also show two more tests in Figure~\ref{fig:sims} to further understand physical drivers behind the observed LF features. Firstly, if we only apply the $10^7\,M_\odot$ mass cutoff for the LRDs while assuming the same underlying Eddington ratio distributions for both LRDs and quasars, LRDs only trace the faint end of a seemingly shared LF between the two populations (shown in the middle row of Figure~\ref{fig:sims}) without an elevated number density as observed \citep{Akins2024, Kokorev2024lrd, Kocevski2025}. This is expected, as less massive black holes are less luminous at a given Eddington ratio, so the mass cutoff would only truncate the full black hole luminosity function into two regimes. Yet, we could go one step further and assume the Eddington ratio distributions of the LRDs and quasars have the same scatter but the one for LRD is shifted to higher values. No sharp LF cutoff is reproduced for the LRDs although their number density is boosted higher than that of the quasars (shown in the bottom row of Figure~\ref{fig:sims}). In this case, elevated Eddington ratios result in less massive objects being scattered into higher-luminosity bins and consequentially boosts LRD number densities at all luminosities. As a result, it seems that the sharp LF cutoff necessitates less scatter in the Eddington ratio distribution, which is a natural result if LRDs are indeed super-Eddington accretors as we discussed previously. 

Therefore, from our heuristic models, we tentatively conclude that LRDs have lower black hole masses (hence the existence of a cutoff), higher Eddington ratios (hence the higher number density at low luminosity), and smaller scatter in the Eddington ratio distribution (hence the steepness of the cutoff) than those of the quasars. Nonetheless, future work constraining all the relevant scaling relations and AGN duty cycles would be required to quantitatively model all these observed LF shapes.

In summary, although the one luminous LRD candidate that we identify would require spectroscopic follow-up to confirm its broad-line nature, by finding only a single luminous candidate in $\sim100\times$ larger survey volume relative to JWST searches robustly confirms the previously suspected luminosity cutoff in the LRD luminosity function. We suspect that LRDs, unlike quasars, must be predominantly near/super-Eddington accretion onto low-mass black holes in order to produce such a LF cutoff at fainter luminosity than quasars do. Finding even brighter LRDs than our current search may further constrain the shape of the LRD luminosity function as well as more easily detecting the weak 200--300\,K dust emission (if any). However, the area required for such a search may only be possible when combining data from near-infrared missions such as \textit{Roman}, \textit{Euclid}, \textit{SPHEREx}, and the Spitzer archive with ground-based optical surveys such as LSST. 

\section*{Acknowledgment} 
YM thanks Hanpu Liu and Minghao Guo for useful discussion. YM is also grateful for the suggestions from Xiaowei Ou and Jiaxuan Li on figure improvement. MA acknowledges support by the National Aeronautics and Space Administration (NASA) through an award (RSA 1628138) issued by JPL/Caltech. DM, LR, and AS acknowledge support by NASA under award number 80NSSC21K0630, issued through the Astrophysics Data Analysis Program (ADAP). 

The Hyper Suprime-Cam (HSC) collaboration includes the astronomical communities of Japan and Taiwan, and Princeton University.  The HSC instrumentation and software were developed by the National Astronomical Observatory of Japan (NAOJ), the Kavli Institute for the Physics and Mathematics of the Universe (Kavli IPMU), the University of Tokyo, the High Energy Accelerator Research Organization (KEK), the Academia Sinica Institute for Astronomy and Astrophysics in Taiwan (ASIAA), and Princeton University.  Funding was contributed by the FIRST program from the Japanese Cabinet Office, the Ministry of Education, Culture, Sports, Science and Technology (MEXT), the Japan Society for the Promotion of Science (JSPS), Japan Science and Technology Agency (JST), the Toray Science Foundation, NAOJ, Kavli IPMU, KEK, ASIAA, and Princeton University.

This paper is based in part on data collected at the Subaru Telescope and retrieved from the HSC data archive system, which is operated by Subaru Telescope and Astronomy Data Center (ADC) at NAOJ. Data analysis was in part carried out with the cooperation of Center for Computational Astrophysics (CfCA) at NAOJ.  We are honored and grateful for the opportunity of observing the Universe from Mauna Kea, which has the cultural, historical and natural significance in Hawai`i.

This paper makes use of software developed for Vera C. Rubin Observatory \citep{Juric2017LSSTdatamanagement, Bosch2019lsstpipe, Ivezic2019lsst}. We thank the Rubin Observatory for making their code available as free software at \href{http://pipelines.lsst.io/}{http://pipelines.lsst.io/}. 

This paper also partially makes use of the CLAUDS data \citep{Sawicki2019clauds}, whose data products can be accessed from \href{https://www.clauds.net}{https://www.clauds.net}.

\restartappendixnumbering
\appendix\label{sec:Appendix}

\section{Detailed Breakdown of the Ground-based LRD Selection}\label{app:selection}
In this appendix, we provide a detailed breakdown of the number of objects satisfying the selection criteria outlined in Section~\ref{sec:selection}. The original \utoirac\ catalog includes a total of 20,485,012 sources. We now apply the quality cuts and LRD selection criteria in sequential orders: 
\begin{enumerate}[start=0]
    \item after applying the quality cuts, 10,503,279 sources remain;
    \item after applying the SNR cuts, 1,913,487 sources remain;
    \item after applying the $g$-band drop-out cut, 20,668 sources remain; 
    \item after applying the $K/K_\mathrm{s}$-band magnitude cut, 15,664 sources remain;
    \item after applying the $i-y$ cut, 1,782 sources remain;
    \item after applying the $K-[3.6]$ cut, 41 sources remain;
    \item after applying the \texttt{extendedness=0} cut, 2 sources remain;
    \item after visual inspection of HST/JWST imaging data for compactness in higher spatial resolution, only one single source---\source---remains. 
\end{enumerate}

\section{Photometric Redshift Estimation of \source}\label{app:eazy}

\begin{figure}[ht!]
    \centering
    \includegraphics[width=0.75\columnwidth]{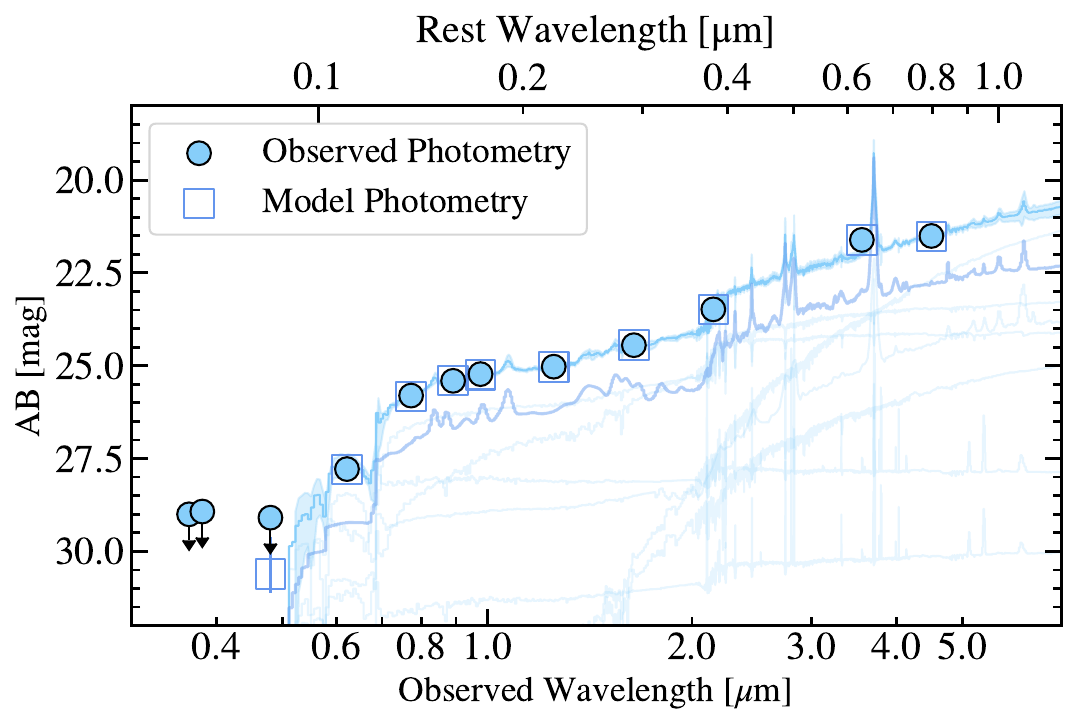}
    \caption{The observed/model photometry of \source\ (circles/squares) is shown on top of the best-fit \texttt{EAZY} model (light blue curve with shades). The sub-components of the best-fit model are shown with smaller line widths. Among all sub-components, UNCOVER-45924, a JWST-selected LRD, is the most dominant component of all and is shown with thicker line width.}
    \label{fig:eazy}
\end{figure}

For the photo-$z$ estimation, following \cite{Ma2025z2lrd}, we add the best-fit models of UNCOVER-45924 \citep{Labbe2024monster} and RUBIES-BLAGN-1 \citep{Wang2025brd} into the \texttt{tweak\_fsps\_QSF\_12\_v3} template suite to include the high-EW H$\alpha$ emission line and the red optical colors not characterized by the default suite. We use a flat redshift prior between $z=0$ and $z=8$ and yield $z_\mathrm{phot}=4.65_{-0.04}^{+0.18}$. The posterior distribution of the redshift is already shown in the inset of Figure~\ref{fig:sed} and fall almost all within the redshift range that we are interested in. 

We also show the best-fit \texttt{EAZY} model and its sub-components in Figure~\ref{fig:eazy}. It is worth noting that in the best-fit \texttt{EAZY} model shown in Figure~\ref{fig:eazy}, the spectrum of UNCOVER-45924 is the most dominant template for its H$\alpha$ emission line and its redness. Although interpreting the physicality of \texttt{EAZY} models should be done with extreme caution, this feature likely alludes to the fact that the true SED shape of \source\ is rather similar to that of spectroscopically confirmed LRDs. 

% $M_{5100}=-23.92_{-0.13}^{+0.09}$ 

\section{Compact Morphology of \source}\label{app:source_morphology}

\begin{figure}[ht]
    \centering
    \includegraphics[width=\columnwidth]{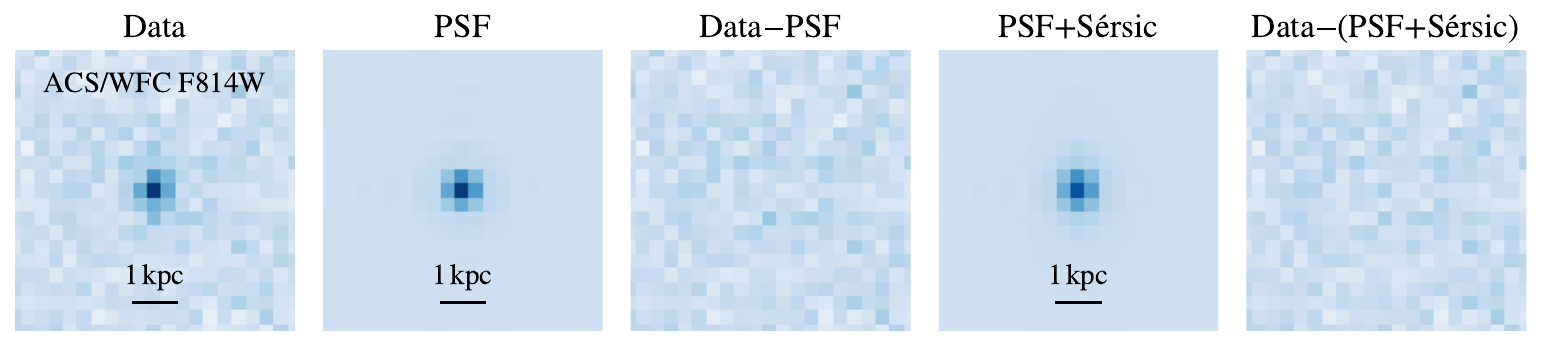}
    \caption{The HST ACS/WFC F814W image cutout of \source, the PSF and the PSF+S\'ersic composite models, as well as their residuals are shown. The cutout and the corresponding point spreading function used in this work are available at \href{https://grizli-cutout.herokuapp.com/}{https://grizli-cutout.herokuapp.com/}.}
    \label{fig:morphology}
\end{figure}

\source\ is within the COSMOS footprints \citep{Koekemoer2007cosmoshst, Scoville2007cosmos} and has ACS/WFC-F814W imaging available from the HST---this filter roughly traces the rest-frame 1490\,\AA\ emission. We show the image cutout in Figure~\ref{fig:morphology}. We use \texttt{pysersic} \citep{Pasha&Miller2023pysersic} to characterize the morphology of the source. The pixel scale of the cutout is $0.''05$/pix, and the point spread function (PSF) is measured to be $0.''10$ in full-width-at-half-maximum (FWHM) near the source position. By visual inspection, \source\ spans roughly three pixels and is consistent with being PSF-dominated. We therefore only fit a point source model onto the F814W image and obtained a residual consistent with the background. As a sanity check, we also add a \cite{Sersic1963} profile into the model and obtain $r_\mathrm{e}\approx0.''025$, which is fully within the pixel scale and the PSF. Therefore, we conclude that \source\ is indeed a compact source with a conservative size upper limit of $r_\mathrm{e}<310\,\mathrm{pc}$ set by the PSF FWHM---this is size is very similar to the LRDs selected with JWST \citep{Akins2024, Kokorev2024lrd, Chen2025host}. Both models and their corresponding residuals are also shown in Figure~\ref{fig:morphology}. 

\bibliography{ref}{}

\begin{thebibliography}{}
\expandafter\ifx\csname natexlab\endcsname\relax\def\natexlab#1{#1}\fi
\providecommand{\url}[1]{\href{#1}{#1}}
\providecommand{\dodoi}[1]{doi:~\href{http://doi.org/#1}{\nolinkurl{#1}}}
\providecommand{\doeprint}[1]{\href{http://ascl.net/#1}{\nolinkurl{http://ascl.net/#1}}}
\providecommand{\doarXiv}[1]{\href{https://arxiv.org/abs/#1}{\nolinkurl{https://arxiv.org/abs/#1}}}

\bibitem[{M.~A. {Abramowicz} {et~al.}(1988){Abramowicz}, {Czerny}, {Lasota}, \& {Szuszkiewicz}}]{Abramowicz1988slimdisk}
{Abramowicz}, M.~A., {Czerny}, B., {Lasota}, J.~P., \& {Szuszkiewicz}, E. 1988, \bibinfo{title}{{Slim Accretion Disks},} \apj, 332, 646, \dodoi{10.1086/166683}

\bibitem[{H. {Aihara} {et~al.}(2018){Aihara}, {Arimoto}, {Armstrong}, {Arnouts}, {Bahcall}, {Bickerton}, {Bosch}, {Bundy}, {Capak}, {Chan}, {Chiba}, {Coupon}, {Egami}, {Enoki}, {Finet}, {Fujimori}, {Fujimoto}, {Furusawa}, {Furusawa}, {Goto}, {Goulding}, {Greco}, {Greene}, {Gunn}, {Hamana}, {Harikane}, {Hashimoto}, {Hattori}, {Hayashi}, {Hayashi}, {He{\l}miniak}, {Higuchi}, {Hikage}, {Ho}, {Hsieh}, {Huang}, {Huang}, {Ikeda}, {Imanishi}, {Inoue}, {Iwasawa}, {Iwata}, {Jaelani}, {Jian}, {Kamata}, {Karoji}, {Kashikawa}, {Katayama}, {Kawanomoto}, {Kayo}, {Koda}, {Koike}, {Kojima}, {Komiyama}, {Konno}, {Koshida}, {Koyama}, {Kusakabe}, {Leauthaud}, {Lee}, {Lin}, {Lin}, {Lupton}, {Mandelbaum}, {Matsuoka}, {Medezinski}, {Mineo}, {Miyama}, {Miyatake}, {Miyazaki}, {Momose}, {More}, {More}, {Moritani}, {Moriya}, {Morokuma}, {Mukae}, {Murata}, {Murayama}, {Nagao}, {Nakata}, {Niida}, {Niikura}, {Nishizawa}, {Obuchi}, {Oguri}, {Oishi}, {Okabe}, {Okamoto}, {Okura}, {Ono}, {Onodera}, {Onoue}, {Osato}, {Ouchi}, {Price}, {Pyo},
  {Sako}, {Sawicki}, {Shibuya}, {Shimasaku}, {Shimono}, {Shirasaki}, {Silverman}, {Simet}, {Speagle}, {Spergel}, {Strauss}, {Sugahara}, {Sugiyama}, {Suto}, {Suyu}, {Suzuki}, {Tait}, {Takada}, {Takata}, {Tamura}, {Tanaka}, {Tanaka}, {Tanaka}, {Tanaka}, {Terai}, {Terashima}, {Toba}, {Tominaga}, {Toshikawa}, {Turner}, {Uchida}, {Uchiyama}, {Umetsu}, {Uraguchi}, {Urata}, {Usuda}, {Utsumi}, {Wang}, {Wang}, {Wong}, {Yabe}, {Yamada}, {Yamanoi}, {Yasuda}, {Yeh}, {Yonehara}, \& {Yuma}}]{Aihara2018hscssp}
{Aihara}, H., {Arimoto}, N., {Armstrong}, R., {et~al.} 2018, \bibinfo{title}{{The Hyper Suprime-Cam SSP Survey: Overview and survey design},} \pasj, 70, S4, \dodoi{10.1093/pasj/psx066}

\bibitem[{H. {Aihara} {et~al.}(2022){Aihara}, {AlSayyad}, {Ando}, {Armstrong}, {Bosch}, {Egami}, {Furusawa}, {Furusawa}, {Harasawa}, {Harikane}, {Hsieh}, {Ikeda}, {Ito}, {Iwata}, {Kodama}, {Koike}, {Kokubo}, {Komiyama}, {Li}, {Liang}, {Lin}, {Lupton}, {Lust}, {MacArthur}, {Mawatari}, {Mineo}, {Miyatake}, {Miyazaki}, {More}, {Morishima}, {Murayama}, {Nakajima}, {Nakata}, {Nishizawa}, {Oguri}, {Okabe}, {Okura}, {Ono}, {Osato}, {Ouchi}, {Pan}, {Plazas Malag{\'o}n}, {Price}, {Reed}, {Rykoff}, {Shibuya}, {Simunovic}, {Strauss}, {Sugimori}, {Suto}, {Suzuki}, {Takada}, {Takagi}, {Takata}, {Takita}, {Tanaka}, {Tang}, {Taranu}, {Terai}, {Toba}, {Turner}, {Uchiyama}, {Vijarnwannaluk}, {Waters}, {Yamada}, {Yamamoto}, \& {Yamashita}}]{Aihara2022hscdr3}
{Aihara}, H., {AlSayyad}, Y., {Ando}, M., {et~al.} 2022, \bibinfo{title}{{Third data release of the Hyper Suprime-Cam Subaru Strategic Program},} \pasj, 74, 247, \dodoi{10.1093/pasj/psab122}

\bibitem[{H.~B. {Akins} {et~al.}(2024){Akins}, {Casey}, {Lambrides}, {Allen}, {Andika}, {Brinch}, {Champagne}, {Cooper}, {Ding}, {Drakos}, {Faisst}, {Finkelstein}, {Franco}, {Fujimoto}, {Gentile}, {Gillman}, {Gozaliasl}, {Harish}, {Hayward}, {Hirschmann}, {Ilbert}, {Kartaltepe}, {Kocevski}, {Koekemoer}, {Kokorev}, {Liu}, {Long}, {McCracken}, {McKinney}, {Onoue}, {Paquereau}, {Renzini}, {Rhodes}, {Robertson}, {Shuntov}, {Silverman}, {Tanaka}, {Toft}, {Trakhtenbrot}, {Valentino}, \& {Zavala}}]{Akins2024}
{Akins}, H.~B., {Casey}, C.~M., {Lambrides}, E., {et~al.} 2024, \bibinfo{title}{{COSMOS-Web: The over-abundance and physical nature of ``little red dots''--Implications for early galaxy and SMBH assembly},} arXiv e-prints, arXiv:2406.10341, \dodoi{10.48550/arXiv.2406.10341}

\bibitem[{T.~T. {Ananna} {et~al.}(2024){Ananna}, {Bogd{\'a}n}, {Kov{\'a}cs}, {Natarajan}, \& {Hickox}}]{Ananna2024}
{Ananna}, T.~T., {Bogd{\'a}n}, {\'A}., {Kov{\'a}cs}, O.~E., {Natarajan}, P., \& {Hickox}, R.~C. 2024, \bibinfo{title}{{X-Ray View of Little Red Dots: Do They Host Supermassive Black Holes?},} \apjl, 969, L18, \dodoi{10.3847/2041-8213/ad5669}

\bibitem[{M. {Annunziatella} {et~al.}(2023){Annunziatella}, {Sajina}, {Stefanon}, {Marchesini}, {Lacy}, {Labb{\'e}}, {Houston}, {Bezanson}, {Egami}, {Fan}, {Farrah}, {Greene}, {Goulding}, {Lin}, {Liu}, {Moutard}, {Ono}, {Ouchi}, {Sawicki}, {Surace}, \& {Whitaker}}]{Annunziatella2023shiraz}
{Annunziatella}, M., {Sajina}, A., {Stefanon}, M., {et~al.} 2023, \bibinfo{title}{{The Spitzer Coverage of HSC-Deep with IRAC for Z studies (SHIRAZ). I. IRAC Mosaics},} \aj, 166, 25, \dodoi{10.3847/1538-3881/acd773}

\bibitem[{J. {Bosch} {et~al.}(2018){Bosch}, {Armstrong}, {Bickerton}, {Furusawa}, {Ikeda}, {Koike}, {Lupton}, {Mineo}, {Price}, {Takata}, {Tanaka}, {Yasuda}, {AlSayyad}, {Becker}, {Coulton}, {Coupon}, {Garmilla}, {Huang}, {Krughoff}, {Lang}, {Leauthaud}, {Lim}, {Lust}, {MacArthur}, {Mandelbaum}, {Miyatake}, {Miyazaki}, {Murata}, {More}, {Okura}, {Owen}, {Swinbank}, {Strauss}, {Yamada}, \& {Yamanoi}}]{Bosch2018hscpipe}
{Bosch}, J., {Armstrong}, R., {Bickerton}, S., {et~al.} 2018, \bibinfo{title}{{The Hyper Suprime-Cam software pipeline},} \pasj, 70, S5, \dodoi{10.1093/pasj/psx080}

\bibitem[{J. {Bosch} {et~al.}(2019){Bosch}, {AlSayyad}, {Armstrong}, {Bellm}, {Chiang}, {Eggl}, {Findeisen}, {Fisher-Levine}, {Guy}, {Guyonnet}, {Ivezi{\'c}}, {Jenness}, {Kov{\'a}cs}, {Krughoff}, {Lupton}, {Lust}, {MacArthur}, {Meyers}, {Moolekamp}, {Morrison}, {Morton}, {O'Mullane}, {Parejko}, {Plazas}, {Price}, {Rawls}, {Reed}, {Schellart}, {Slater}, {Sullivan}, {Swinbank}, {Taranu}, {Waters}, \& {Wood-Vasey}}]{Bosch2019lsstpipe}
{Bosch}, J., {AlSayyad}, Y., {Armstrong}, R., {et~al.} 2019, in Astronomical Society of the Pacific Conference Series, Vol. 523, Astronomical Data Analysis Software and Systems XXVII, ed. P.~J. {Teuben}, M.~W. {Pound}, B.~A. {Thomas}, \& E.~M. {Warner}, 521, \dodoi{10.48550/arXiv.1812.03248}

\bibitem[{B.~J. {Boyle} {et~al.}(1988){Boyle}, {Shanks}, \& {Peterson}}]{Boyle1988}
{Boyle}, B.~J., {Shanks}, T., \& {Peterson}, B.~A. 1988, \bibinfo{title}{{The evolution of optically selected QSOs - II.},} \mnras, 235, 935, \dodoi{10.1093/mnras/235.3.935}

\bibitem[{G.~B. {Brammer} {et~al.}(2008){Brammer}, {van Dokkum}, \& {Coppi}}]{Brammer2008eazy}
{Brammer}, G.~B., {van Dokkum}, P.~G., \& {Coppi}, P. 2008, \bibinfo{title}{{EAZY: A Fast, Public Photometric Redshift Code},} \apj, 686, 1503, \dodoi{10.1086/591786}

\bibitem[{A.~J. {Burgasser} {et~al.}(2024){Burgasser}, {Bezanson}, {Labbe}, {Brammer}, {Cutler}, {Furtak}, {Greene}, {Gerasimov}, {Leja}, {Pan}, {Price}, {Wang}, {Weaver}, {Whitaker}, {Fujimoto}, {Kokorev}, {Dayal}, {Nanayakkara}, {Williams}, {Marchesini}, {Zitrin}, \& {van Dokkum}}]{Burgasser2024}
{Burgasser}, A.~J., {Bezanson}, R., {Labbe}, I., {et~al.} 2024, \bibinfo{title}{{UNCOVER: JWST Spectroscopy of Three Cold Brown Dwarfs at Kiloparsec-scale Distances},} \apj, 962, 177, \dodoi{10.3847/1538-4357/ad206f}

\bibitem[{C.~M. {Casey} {et~al.}(2023){Casey}, {Kartaltepe}, {Drakos}, {Franco}, {Harish}, {Paquereau}, {Ilbert}, {Rose}, {Cox}, {Nightingale}, {Robertson}, {Silverman}, {Koekemoer}, {Massey}, {McCracken}, {Rhodes}, {Akins}, {Allen}, {Amvrosiadis}, {Arango-Toro}, {Bagley}, {Bongiorno}, {Capak}, {Champagne}, {Chartab}, {Ch{\'a}vez Ortiz}, {Chworowsky}, {Cooke}, {Cooper}, {Darvish}, {Ding}, {Faisst}, {Finkelstein}, {Fujimoto}, {Gentile}, {Gillman}, {Gould}, {Gozaliasl}, {Hayward}, {He}, {Hemmati}, {Hirschmann}, {Jahnke}, {Jin}, {Khostovan}, {Kokorev}, {Lambrides}, {Laigle}, {Larson}, {Leung}, {Liu}, {Liaudat}, {Long}, {Magdis}, {Mahler}, {Mainieri}, {Manning}, {Maraston}, {Martin}, {McCleary}, {McKinney}, {McPartland}, {Mobasher}, {Pattnaik}, {Renzini}, {Rich}, {Sanders}, {Sattari}, {Scognamiglio}, {Scoville}, {Sheth}, {Shuntov}, {Sparre}, {Suzuki}, {Talia}, {Toft}, {Trakhtenbrot}, {Urry}, {Valentino}, {Vanderhoof}, {Vardoulaki}, {Weaver}, {Whitaker}, {Wilkins}, {Yang}, \& {Zavala}}]{Casey2023cosmosweb}
{Casey}, C.~M., {Kartaltepe}, J.~S., {Drakos}, N.~E., {et~al.} 2023, \bibinfo{title}{{COSMOS-Web: An Overview of the JWST Cosmic Origins Survey},} \apj, 954, 31, \dodoi{10.3847/1538-4357/acc2bc}

\bibitem[{C.~M. {Casey} {et~al.}(2025){Casey}, {Akins}, {Finkelstein}, {Franco}, {Fujimoto}, {Liu}, {Long}, {Magdis}, {Manning}, {McKinney}, {Shuntov}, \& {Tanaka}}]{Casey2025alma}
{Casey}, C.~M., {Akins}, H.~B., {Finkelstein}, S.~L., {et~al.} 2025, \bibinfo{title}{{An upper limit of 10$^6$ M$_\odot$ in dust from ALMA observations in 60 Little Red Dots},} arXiv e-prints, arXiv:2505.18873, \dodoi{10.48550/arXiv.2505.18873}

\bibitem[{C.-H. {Chen} {et~al.}(2025{\natexlab{a}}){Chen}, {Ho}, {Li}, \& {Inayoshi}}]{Chen2025nebular}
{Chen}, C.-H., {Ho}, L.~C., {Li}, R., \& {Inayoshi}, K. 2025{\natexlab{a}}, \bibinfo{title}{{The Physical Nature of the Off-centered Extended Emission Associated with the Little Red Dots},} arXiv e-prints, arXiv:2505.03183, \dodoi{10.48550/arXiv.2505.03183}

\bibitem[{C.-H. {Chen} {et~al.}(2025{\natexlab{b}}){Chen}, {Ho}, {Li}, \& {Zhuang}}]{Chen2025host}
{Chen}, C.-H., {Ho}, L.~C., {Li}, R., \& {Zhuang}, M.-Y. 2025{\natexlab{b}}, \bibinfo{title}{{The Host Galaxy (If Any) of the Little Red Dots},} \apj, 983, 60, \dodoi{10.3847/1538-4357/ada93a}

\bibitem[{P. {Dayal} \& S.~K. {Giri}(2024){Dayal} \& {Giri}}]{Dayal&Giri2024}
{Dayal}, P., \& {Giri}, S.~K. 2024, \bibinfo{title}{{Warm dark matter constraints from the JWST},} \mnras, 528, 2784, \dodoi{10.1093/mnras/stae176}

\bibitem[{A. {de Graaff} {et~al.}(2025){de Graaff}, {Rix}, {Naidu}, {Labbe}, {Wang}, {Leja}, {Matthee}, {Katz}, {Greene}, {Hviding}, {Baggen}, {Bezanson}, {Boogaard}, {Brammer}, {Dayal}, {van Dokkum}, {Goulding}, {Hirschmann}, {Maseda}, {McConachie}, {Miller}, {Nelson}, {Oesch}, {Setton}, {Shivaei}, {Weibel}, {Whitaker}, \& {Williams}}]{deGraaff2025cliff}
{de Graaff}, A., {Rix}, H.-W., {Naidu}, R.~P., {et~al.} 2025, \bibinfo{title}{{A remarkable Ruby: Absorption in dense gas, rather than evolved stars, drives the extreme Balmer break of a Little Red Dot at $z=3.5$},} arXiv e-prints, arXiv:2503.16600, \dodoi{10.48550/arXiv.2503.16600}

\bibitem[{G. {Desprez} {et~al.}(2023){Desprez}, {Picouet}, {Moutard}, {Arnouts}, {Sawicki}, {Coupon}, {Gwyn}, {Chen}, {Huang}, {Golob}, {Furusawa}, {Ikeda}, {Paltani}, {Cheng}, {Hartley}, {Hsieh}, {Ilbert}, {Kauffmann}, {McCracken}, {Shuntov}, {Tanaka}, {Toft}, {Tresse}, \& {Weaver}}]{Desprez2023clauds+hsc}
{Desprez}, G., {Picouet}, V., {Moutard}, T., {et~al.} 2023, \bibinfo{title}{{Combining the CLAUDS and HSC-SSP surveys. U + grizy(+YJHK$_{s}$) photometry and photometric redshifts for 18M galaxies in the 20 deg$^{2}$ of the HSC-SSP Deep and ultraDeep fields},} \aap, 670, A82, \dodoi{10.1051/0004-6361/202243363}

\bibitem[{L.~J. {Furtak} {et~al.}(2024){Furtak}, {Labb{\'e}}, {Zitrin}, {Greene}, {Dayal}, {Chemerynska}, {Kokorev}, {Miller}, {Goulding}, {de Graaff}, {Bezanson}, {Brammer}, {Cutler}, {Leja}, {Pan}, {Price}, {Wang}, {Weaver}, {Whitaker}, {Atek}, {Bogd{\'a}n}, {Charlot}, {Curtis-Lake}, {van Dokkum}, {Endsley}, {Feldmann}, {Fudamoto}, {Fujimoto}, {Glazebrook}, {Juneau}, {Marchesini}, {Maseda}, {Nelson}, {Oesch}, {Plat}, {Setton}, {Stark}, \& {Williams}}]{Furtak2024qso1}
{Furtak}, L.~J., {Labb{\'e}}, I., {Zitrin}, A., {et~al.} 2024, \bibinfo{title}{{A high black-hole-to-host mass ratio in a lensed AGN in the early Universe},} \nat, 628, 57, \dodoi{10.1038/s41586-024-07184-8}

\bibitem[{N. {Gehrels}(1986){Gehrels}}]{Gehrels1986}
{Gehrels}, N. 1986, \bibinfo{title}{{Confidence Limits for Small Numbers of Events in Astrophysical Data},} \apj, 303, 336, \dodoi{10.1086/164079}

\bibitem[{J.~E. {Greene} {et~al.}(2020){Greene}, {Strader}, \& {Ho}}]{Greene2020}
{Greene}, J.~E., {Strader}, J., \& {Ho}, L.~C. 2020, \bibinfo{title}{{Intermediate-Mass Black Holes},} \araa, 58, 257, \dodoi{10.1146/annurev-astro-032620-021835}

\bibitem[{J.~E. {Greene} {et~al.}(2024){Greene}, {Labbe}, {Goulding}, {Furtak}, {Chemerynska}, {Kokorev}, {Dayal}, {Volonteri}, {Williams}, {Wang}, {Setton}, {Burgasser}, {Bezanson}, {Atek}, {Brammer}, {Cutler}, {Feldmann}, {Fujimoto}, {Glazebrook}, {de Graaff}, {Khullar}, {Leja}, {Marchesini}, {Maseda}, {Matthee}, {Miller}, {Naidu}, {Nanayakkara}, {Oesch}, {Pan}, {Papovich}, {Price}, {van Dokkum}, {Weaver}, {Whitaker}, \& {Zitrin}}]{Greene2024}
{Greene}, J.~E., {Labbe}, I., {Goulding}, A.~D., {et~al.} 2024, \bibinfo{title}{{UNCOVER Spectroscopy Confirms the Surprising Ubiquity of Active Galactic Nuclei in Red Sources at z > 5},} \apj, 964, 39, \dodoi{10.3847/1538-4357/ad1e5f}

\bibitem[{W. {He} {et~al.}(2024){He}, {Akiyama}, {Enoki}, {Ichikawa}, {Inayoshi}, {Kashikawa}, {Kawaguchi}, {Matsuoka}, {Nagao}, {Onoue}, {Oogi}, {Schulze}, {Toba}, \& {Ueda}}]{He2024hscqsoz4}
{He}, W., {Akiyama}, M., {Enoki}, M., {et~al.} 2024, \bibinfo{title}{{Black Hole Mass and Eddington-ratio Distributions of Less-luminous Quasars at z {\ensuremath{\sim}} 4 in the Subaru Hyper Suprime-Cam Wide Field},} \apj, 962, 152, \dodoi{10.3847/1538-4357/ad1518}

\bibitem[{R.~E. {Hviding} {et~al.}(2025){Hviding}, {de Graaff}, {Miller}, {Setton}, {Greene}, {Labb{\'e}}, {Brammer}, {Bezanson}, {Boogaard}, {Cleri}, {Leja}, {Maseda}, {McConachie}, {Matthee}, {Naidu}, {Oesch}, {Wang}, {Whitaker}, \& {Williams}}]{Hviding2025}
{Hviding}, R.~E., {de Graaff}, A., {Miller}, T.~B., {et~al.} 2025, \bibinfo{title}{{RUBIES: A Spectroscopic Census of Little Red Dots; All V-Shaped Point Sources Have Broad Lines},} arXiv e-prints, arXiv:2506.05459, \dodoi{10.48550/arXiv.2506.05459}

\bibitem[{K. {Inayoshi}(2025){Inayoshi}}]{Inayoshi2025lognormal}
{Inayoshi}, K. 2025, \bibinfo{title}{{Little Red Dots as the Very First Activity of Black Hole Growth},} \apjl, 988, L22, \dodoi{10.3847/2041-8213/adea66}

\bibitem[{K. {Inayoshi} {et~al.}(2024){Inayoshi}, {Kimura}, \& {Noda}}]{Inayoshi2024superEdd}
{Inayoshi}, K., {Kimura}, S.~S., \& {Noda}, H. 2024, \bibinfo{title}{{Weakness of X-rays and Variability in High-redshift AGNs with Super-Eddington Accretion},} arXiv e-prints, arXiv:2412.03653, \dodoi{10.48550/arXiv.2412.03653}

\bibitem[{K. {Inayoshi} \& R. {Maiolino}(2025){Inayoshi} \& {Maiolino}}]{Inayoshi&Maiolino2025}
{Inayoshi}, K., \& {Maiolino}, R. 2025, \bibinfo{title}{{Extremely Dense Gas around Little Red Dots and High-redshift Active Galactic Nuclei: A Nonstellar Origin of the Balmer Break and Absorption Features},} \apjl, 980, L27, \dodoi{10.3847/2041-8213/adaebd}

\bibitem[{K. {Inayoshi} {et~al.}(2020){Inayoshi}, {Visbal}, \& {Haiman}}]{Inayoshi2020araa}
{Inayoshi}, K., {Visbal}, E., \& {Haiman}, Z. 2020, \bibinfo{title}{{The Assembly of the First Massive Black Holes},} \araa, 58, 27, \dodoi{10.1146/annurev-astro-120419-014455}

\bibitem[{{\v{Z}}. {Ivezi{\'c}} {et~al.}(2019){Ivezi{\'c}}, {Kahn}, {Tyson}, {Abel}, {Acosta}, {Allsman}, {Alonso}, {AlSayyad}, {Anderson}, {Andrew}, {Angel}, {Angeli}, {Ansari}, {Antilogus}, {Araujo}, {Armstrong}, {Arndt}, {Astier}, {Aubourg}, {Auza}, {Axelrod}, {Bard}, {Barr}, {Barrau}, {Bartlett}, {Bauer}, {Bauman}, {Baumont}, {Bechtol}, {Bechtol}, {Becker}, {Becla}, {Beldica}, {Bellavia}, {Bianco}, {Biswas}, {Blanc}, {Blazek}, {Blandford}, {Bloom}, {Bogart}, {Bond}, {Booth}, {Borgland}, {Borne}, {Bosch}, {Boutigny}, {Brackett}, {Bradshaw}, {Brandt}, {Brown}, {Bullock}, {Burchat}, {Burke}, {Cagnoli}, {Calabrese}, {Callahan}, {Callen}, {Carlin}, {Carlson}, {Chandrasekharan}, {Charles-Emerson}, {Chesley}, {Cheu}, {Chiang}, {Chiang}, {Chirino}, {Chow}, {Ciardi}, {Claver}, {Cohen-Tanugi}, {Cockrum}, {Coles}, {Connolly}, {Cook}, {Cooray}, {Covey}, {Cribbs}, {Cui}, {Cutri}, {Daly}, {Daniel}, {Daruich}, {Daubard}, {Daues}, {Dawson}, {Delgado}, {Dellapenna}, {de Peyster}, {de Val-Borro}, {Digel}, {Doherty},
  {Dubois}, {Dubois-Felsmann}, {Durech}, {Economou}, {Eifler}, {Eracleous}, {Emmons}, {Fausti Neto}, {Ferguson}, {Figueroa}, {Fisher-Levine}, {Focke}, {Foss}, {Frank}, {Freemon}, {Gangler}, {Gawiser}, {Geary}, {Gee}, {Geha}, {Gessner}, {Gibson}, {Gilmore}, {Glanzman}, {Glick}, {Goldina}, {Goldstein}, {Goodenow}, {Graham}, {Gressler}, {Gris}, {Guy}, {Guyonnet}, {Haller}, {Harris}, {Hascall}, {Haupt}, {Hernandez}, {Herrmann}, {Hileman}, {Hoblitt}, {Hodgson}, {Hogan}, {Howard}, {Huang}, {Huffer}, {Ingraham}, {Innes}, {Jacoby}, {Jain}, {Jammes}, {Jee}, {Jenness}, {Jernigan}, {Jevremovi{\'c}}, {Johns}, {Johnson}, {Johnson}, {Jones}, {Juramy-Gilles}, {Juri{\'c}}, {Kalirai}, {Kallivayalil}, {Kalmbach}, {Kantor}, {Karst}, {Kasliwal}, {Kelly}, {Kessler}, {Kinnison}, {Kirkby}, {Knox}, {Kotov}, {Krabbendam}, {Krughoff}, {Kub{\'a}nek}, {Kuczewski}, {Kulkarni}, {Ku}, {Kurita}, {Lage}, {Lambert}, {Lange}, {Langton}, {Le Guillou}, {Levine}, {Liang}, {Lim}, {Lintott}, {Long}, {Lopez}, {Lotz}, {Lupton}, {Lust}, {MacArthur},
  {Mahabal}, {Mandelbaum}, {Markiewicz}, {Marsh}, {Marshall}, {Marshall}, {May}, {McKercher}, {McQueen}, {Meyers}, {Migliore}, {Miller}, \& {Mills}}]{Ivezic2019lsst}
{Ivezi{\'c}}, {\v{Z}}., {Kahn}, S.~M., {Tyson}, J.~A., {et~al.} 2019, \bibinfo{title}{{LSST: From Science Drivers to Reference Design and Anticipated Data Products},} \apj, 873, 111, \dodoi{10.3847/1538-4357/ab042c}

\bibitem[{M.~J. {Jarvis} {et~al.}(2013){Jarvis}, {Bonfield}, {Bruce}, {Geach}, {McAlpine}, {McLure}, {Gonz{\'a}lez-Solares}, {Irwin}, {Lewis}, {Yoldas}, {Andreon}, {Cross}, {Emerson}, {Dalton}, {Dunlop}, {Hodgkin}, {Le}, {Karouzos}, {Meisenheimer}, {Oliver}, {Rawlings}, {Simpson}, {Smail}, {Smith}, {Sullivan}, {Sutherland}, {White}, \& {Zwart}}]{Jarvis2013video}
{Jarvis}, M.~J., {Bonfield}, D.~G., {Bruce}, V.~A., {et~al.} 2013, \bibinfo{title}{{The VISTA Deep Extragalactic Observations (VIDEO) survey},} \mnras, 428, 1281, \dodoi{10.1093/mnras/sts118}

\bibitem[{X. {Ji} {et~al.}(2025{\natexlab{a}}){Ji}, {Maiolino}, {{\"U}bler}, {Scholtz}, {D'Eugenio}, {Sun}, {Perna}, {Turner}, {Arribas}, {Bennett}, {Bunker}, {Carniani}, {Charlot}, {Cresci}, {Curti}, {Egami}, {Fabian}, {Inayoshi}, {Isobe}, {Jones}, {Juod{\v{z}}balis}, {Kumari}, {Lyu}, {Mazzolari}, {Parlanti}, {Robertson}, {Rodr{\'\i}guez Del Pino}, {Schneider}, {Sijacki}, {Tacchella}, {Trinca}, {Valiante}, {Venturi}, {Volonteri}, {Willott}, {Witten}, \& {Witstok}}]{Ji2025blackthunder}
{Ji}, X., {Maiolino}, R., {{\"U}bler}, H., {et~al.} 2025{\natexlab{a}}, \bibinfo{title}{{BlackTHUNDER -- A non-stellar Balmer break in a black hole-dominated little red dot at $z=7.04$},} arXiv e-prints, arXiv:2501.13082, \dodoi{10.48550/arXiv.2501.13082}

\bibitem[{X. {Ji} {et~al.}(2025{\natexlab{b}}){Ji}, {D'Eugenio}, {Juod{\v{z}}balis}, {Walton}, {Fabian}, {Maiolino}, {Ramos Almeida}, {Acosta Pulido}, {Belokurov}, {Isobe}, {Jones}, {Maraston}, {Scholtz}, {Simmonds}, {Tacchella}, {Terlevich}, \& {Terlevich}}]{Ji2025local}
{Ji}, X., {D'Eugenio}, F., {Juod{\v{z}}balis}, I., {et~al.} 2025{\natexlab{b}}, \bibinfo{title}{{Lord of LRDs: Insights into a ``Little Red Dot'' with a low-ionization spectrum at z = 0.1},} arXiv e-prints, arXiv:2507.23774, \dodoi{10.48550/arXiv.2507.23774}

\bibitem[{Y.-F. {Jiang} {et~al.}(2014){Jiang}, {Stone}, \& {Davis}}]{Jiang2014}
{Jiang}, Y.-F., {Stone}, J.~M., \& {Davis}, S.~W. 2014, \bibinfo{title}{{A Global Three-dimensional Radiation Magneto-hydrodynamic Simulation of Super-Eddington Accretion Disks},} \apj, 796, 106, \dodoi{10.1088/0004-637X/796/2/106}

\bibitem[{I. {Juod{\v{z}}balis} {et~al.}(2024){Juod{\v{z}}balis}, {Ji}, {Maiolino}, {D'Eugenio}, {Scholtz}, {Risaliti}, {Fabian}, {Mazzolari}, {Gilli}, {Prandoni}, {Arribas}, {Bunker}, {Carniani}, {Charlot}, {Curtis-Lake}, {de Graaff}, {Hainline}, {Parlanti}, {Perna}, {P{\'e}rez-Gonz{\'a}lez}, {Robertson}, {Tacchella}, {{\"U}bler}, {Williams}, {Willott}, \& {Witstok}}]{Juodzbalis2024rosetta}
{Juod{\v{z}}balis}, I., {Ji}, X., {Maiolino}, R., {et~al.} 2024, \bibinfo{title}{{JADES - the Rosetta stone of JWST-discovered AGN: deciphering the intriguing nature of early AGN},} \mnras, 535, 853, \dodoi{10.1093/mnras/stae2367}

\bibitem[{M. {Juri{\'c}} {et~al.}(2017){Juri{\'c}}, {Kantor}, {Lim}, {Lupton}, {Dubois-Felsmann}, {Jenness}, {Axelrod}, {Aleksi{\'c}}, {Allsman}, {AlSayyad}, {Alt}, {Armstrong}, {Basney}, {Becker}, {Becla}, {Biswas}, {Bosch}, {Boutigny}, {Kind}, {Ciardi}, {Connolly}, {Daniel}, {Daues}, {Economou}, {Chiang}, {Fausti}, {Fisher-Levine}, {Freemon}, {Gris}, {Hernandez}, {Hoblitt}, {Ivezi{\'c}}, {Jammes}, {Jevremovi{\'c}}, {Jones}, {Kalmbach}, {Kasliwal}, {Krughoff}, {Lurie}, {Lust}, {MacArthur}, {Melchior}, {Moeyens}, {Nidever}, {Owen}, {Parejko}, {Peterson}, {Petravick}, {Pietrowicz}, {Price}, {Reiss}, {Shaw}, {Sick}, {Slater}, {Strauss}, {Sullivan}, {Swinbank}, {Van Dyk}, {Vuj{\v{c}}i{\'c}}, {Withers}, \& {Yoachim}}]{Juric2017LSSTdatamanagement}
{Juri{\'c}}, M., {Kantor}, J., {Lim}, K.~T., {et~al.} 2017, in Astronomical Society of the Pacific Conference Series, Vol. 512, Astronomical Data Analysis Software and Systems XXV, ed. N.~P.~F. {Lorente}, K.~{Shortridge}, \& R.~{Wayth}, 279, \dodoi{10.48550/arXiv.1512.07914}

\bibitem[{D. {Kido} {et~al.}(2025){Kido}, {Ioka}, {Hotokezaka}, {Inayoshi}, \& {Irwin}}]{Kido2025}
{Kido}, D., {Ioka}, K., {Hotokezaka}, K., {Inayoshi}, K., \& {Irwin}, C.~M. 2025, \bibinfo{title}{{Black Hole Envelopes in Little Red Dots},} arXiv e-prints, arXiv:2505.06965, \dodoi{10.48550/arXiv.2505.06965}

\bibitem[{D.~D. {Kocevski} {et~al.}(2023){Kocevski}, {Onoue}, {Inayoshi}, {Trump}, {Arrabal Haro}, {Grazian}, {Dickinson}, {Finkelstein}, {Kartaltepe}, {Hirschmann}, {Aird}, {Holwerda}, {Fujimoto}, {Juneau}, {Amor{\'\i}n}, {Backhaus}, {Bagley}, {Barro}, {Bell}, {Bisigello}, {Calabr{\`o}}, {Cleri}, {Cooper}, {Ding}, {Grogin}, {Ho}, {Hutchison}, {Inoue}, {Jiang}, {Jones}, {Koekemoer}, {Li}, {Li}, {McGrath}, {Molina}, {Papovich}, {P{\'e}rez-Gonz{\'a}lez}, {Pirzkal}, {Wilkins}, {Yang}, \& {Yung}}]{Kocevski2023}
{Kocevski}, D.~D., {Onoue}, M., {Inayoshi}, K., {et~al.} 2023, \bibinfo{title}{{Hidden Little Monsters: Spectroscopic Identification of Low-mass, Broad-line AGNs at z > 5 with CEERS},} \apjl, 954, L4, \dodoi{10.3847/2041-8213/ace5a0}

\bibitem[{D.~D. {Kocevski} {et~al.}(2025){Kocevski}, {Finkelstein}, {Barro}, {Taylor}, {Calabr{\`o}}, {Laloux}, {Buchner}, {Trump}, {Leung}, {Yang}, {Dickinson}, {P{\'e}rez-Gonz{\'a}lez}, {Pacucci}, {Inayoshi}, {Somerville}, {McGrath}, {Akins}, {Bagley}, {Bowler}, {Bisigello}, {Carnall}, {Casey}, {Cheng}, {Cleri}, {Costantin}, {Cullen}, {Davis}, {Donnan}, {Dunlop}, {Ellis}, {Ferguson}, {Fujimoto}, {Fontana}, {Giavalisco}, {Grazian}, {Grogin}, {Hathi}, {Hirschmann}, {Huertas-Company}, {Holwerda}, {Illingworth}, {Juneau}, {Kartaltepe}, {Koekemoer}, {Li}, {Lucas}, {Magee}, {Mason}, {McLeod}, {McLure}, {Napolitano}, {Papovich}, {Pirzkal}, {Rodighiero}, {Santini}, {Wilkins}, \& {Yung}}]{Kocevski2025}
{Kocevski}, D.~D., {Finkelstein}, S.~L., {Barro}, G., {et~al.} 2025, \bibinfo{title}{{The Rise of Faint, Red Active Galactic Nuclei at z > 4: A Sample of Little Red Dots in the JWST Extragalactic Legacy Fields},} \apj, 986, 126, \dodoi{10.3847/1538-4357/adbc7d}

\bibitem[{A.~M. {Koekemoer} {et~al.}(2007){Koekemoer}, {Aussel}, {Calzetti}, {Capak}, {Giavalisco}, {Kneib}, {Leauthaud}, {Le F{\`e}vre}, {McCracken}, {Massey}, {Mobasher}, {Rhodes}, {Scoville}, \& {Shopbell}}]{Koekemoer2007cosmoshst}
{Koekemoer}, A.~M., {Aussel}, H., {Calzetti}, D., {et~al.} 2007, \bibinfo{title}{{The COSMOS Survey: Hubble Space Telescope Advanced Camera for Surveys Observations and Data Processing},} \apjs, 172, 196, \dodoi{10.1086/520086}

\bibitem[{V. {Kokorev} {et~al.}(2024){Kokorev}, {Caputi}, {Greene}, {Dayal}, {Trebitsch}, {Cutler}, {Fujimoto}, {Labb{\'e}}, {Miller}, {Iani}, {Navarro-Carrera}, \& {Rinaldi}}]{Kokorev2024lrd}
{Kokorev}, V., {Caputi}, K.~I., {Greene}, J.~E., {et~al.} 2024, \bibinfo{title}{{A Census of Photometrically Selected Little Red Dots at 4 < z < 9 in JWST Blank Fields},} \apj, 968, 38, \dodoi{10.3847/1538-4357/ad4265}

\bibitem[{G. {Kulkarni} {et~al.}(2019){Kulkarni}, {Worseck}, \& {Hennawi}}]{Kulkarni2019QLF}
{Kulkarni}, G., {Worseck}, G., \& {Hennawi}, J.~F. 2019, \bibinfo{title}{{Evolution of the AGN UV luminosity function from redshift 7.5},} \mnras, 488, 1035, \dodoi{10.1093/mnras/stz1493}

\bibitem[{I. {Labbe} {et~al.}(2024){Labbe}, {Greene}, {Matthee}, {Treiber}, {Kokorev}, {Miller}, {Kramarenko}, {Setton}, {Ma}, {Goulding}, {Bezanson}, {Naidu}, {Williams}, {Atek}, {Brammer}, {Cutler}, {Chemerynska}, {Cloonan}, {Dayal}, {de Graaff}, {Fudamoto}, {Fujimoto}, {Furtak}, {Glazebrook}, {Heintz}, {Leja}, {Marchesini}, {Nanayakkara}, {Nelson}, {Oesch}, {Pan}, {Price}, {Shivaei}, {Sobral}, {Suess}, {van Dokkum}, {Wang}, {Weaver}, {Whitaker}, \& {Zitrin}}]{Labbe2024monster}
{Labbe}, I., {Greene}, J.~E., {Matthee}, J., {et~al.} 2024, \bibinfo{title}{{An unambiguous AGN and a Balmer break in an Ultraluminous Little Red Dot at z=4.47 from Ultradeep UNCOVER and All the Little Things Spectroscopy},} arXiv e-prints, arXiv:2412.04557, \dodoi{10.48550/arXiv.2412.04557}

\bibitem[{I. {Labbe} {et~al.}(2025){Labbe}, {Greene}, {Bezanson}, {Fujimoto}, {Furtak}, {Goulding}, {Matthee}, {Naidu}, {Oesch}, {Atek}, {Brammer}, {Chemerynska}, {Coe}, {Cutler}, {Dayal}, {Feldmann}, {Franx}, {Glazebrook}, {Leja}, {Maseda}, {Marchesini}, {Nanayakkara}, {Nelson}, {Pan}, {Papovich}, {Price}, {Suess}, {Wang}, {Weaver}, {Whitaker}, {Williams}, \& {Zitrin}}]{Labbe2025}
{Labbe}, I., {Greene}, J.~E., {Bezanson}, R., {et~al.} 2025, \bibinfo{title}{{UNCOVER: Candidate Red Active Galactic Nuclei at 3 < z < 7 with JWST and ALMA},} \apj, 978, 92, \dodoi{10.3847/1538-4357/ad3551}

\bibitem[{M. {Lacy} {et~al.}(2021){Lacy}, {Surace}, {Farrah}, {Nyland}, {Afonso}, {Brandt}, {Clements}, {Lagos}, {Maraston}, {Pforr}, {Sajina}, {Sako}, {Vaccari}, {Wilson}, {Ballantyne}, {Barkhouse}, {Brunner}, {Cane}, {Clarke}, {Cooper}, {Cooray}, {Covone}, {D'Andrea}, {Evrard}, {Ferguson}, {Frieman}, {Gonzalez-Perez}, {Gupta}, {Hatziminaoglou}, {Huang}, {Jagannathan}, {Jarvis}, {Jones}, {Kimball}, {Lidman}, {Lubin}, {Marchetti}, {Martini}, {McMahon}, {Mei}, {Messias}, {Murphy}, {Newman}, {Nichol}, {Norris}, {Oliver}, {Perez-Fournon}, {Peters}, {Pierre}, {Polisensky}, {Richards}, {Ridgway}, {R{\"o}ttgering}, {Seymour}, {Shirley}, {Somerville}, {Strauss}, {Suntzeff}, {Thorman}, {van Kampen}, {Verma}, {Wechsler}, \& {Wood-Vasey}}]{Lacy2021deepdrill}
{Lacy}, M., {Surace}, J.~A., {Farrah}, D., {et~al.} 2021, \bibinfo{title}{{A Spitzer survey of Deep Drilling Fields to be targeted by the Vera C. Rubin Observatory Legacy Survey of Space and Time},} \mnras, 501, 892, \dodoi{10.1093/mnras/staa3714}

\bibitem[{E. {Lambrides} {et~al.}(2024){Lambrides}, {Garofali}, {Larson}, {Ptak}, {Chiaberge}, {Long}, {Hutchison}, {Norman}, {McKinney}, {Akins}, {Berg}, {Chisholm}, {Civano}, {Cloonan}, {Endsley}, {Faisst}, {Gilli}, {Gillman}, {Hirschmann}, {Kartaltepe}, {Kocevski}, {Kokorev}, {Pacucci}, {Richardson}, {Stiavelli}, \& {Whalen}}]{Lambrides2024}
{Lambrides}, E., {Garofali}, K., {Larson}, R., {et~al.} 2024, \bibinfo{title}{{The Case for Super-Eddington Accretion: Connecting Weak X-ray and UV Line Emission in JWST Broad-Line AGN During the First Gyr of Cosmic Time},} arXiv e-prints, arXiv:2409.13047, \dodoi{10.48550/arXiv.2409.13047}

\bibitem[{A. {Lawrence} {et~al.}(2007){Lawrence}, {Warren}, {Almaini}, {Edge}, {Hambly}, {Jameson}, {Lucas}, {Casali}, {Adamson}, {Dye}, {Emerson}, {Foucaud}, {Hewett}, {Hirst}, {Hodgkin}, {Irwin}, {Lodieu}, {McMahon}, {Simpson}, {Smail}, {Mortlock}, \& {Folger}}]{Lawrence2007ukidss}
{Lawrence}, A., {Warren}, S.~J., {Almaini}, O., {et~al.} 2007, \bibinfo{title}{{The UKIRT Infrared Deep Sky Survey (UKIDSS)},} \mnras, 379, 1599, \dodoi{10.1111/j.1365-2966.2007.12040.x}

\bibitem[{G.~C.~K. {Leung} {et~al.}(2024){Leung}, {Finkelstein}, {P{\'e}rez-Gonz{\'a}lez}, {Morales}, {Taylor}, {Barro}, {Kocevski}, {Akins}, {Carnall}, {Ch{\'a}vez Ortiz}, {Cleri}, {Cullen}, {Donnan}, {Dunlop}, {Ellis}, {Grogin}, {Hirschmann}, {Koekemoer}, {Kokorev}, {Lucas}, {McLeod}, {Papovich}, \& {Yung}}]{Leung2024miri}
{Leung}, G. C.~K., {Finkelstein}, S.~L., {P{\'e}rez-Gonz{\'a}lez}, P.~G., {et~al.} 2024, \bibinfo{title}{{Exploring the Nature of Little Red Dots: Constraints on AGN and Stellar Contributions from PRIMER MIRI Imaging},} arXiv e-prints, arXiv:2411.12005, \dodoi{10.48550/arXiv.2411.12005}

\bibitem[{W. {Li} {et~al.}(2024){Li}, {Inayoshi}, {Onoue}, {He}, {Matsuoka}, {Pan}, {Akiyama}, {Izumi}, \& {Nagao}}]{Li2024fedd}
{Li}, W., {Inayoshi}, K., {Onoue}, M., {et~al.} 2024, \bibinfo{title}{{Reconstruction of Cosmic Black Hole Growth and Mass Distribution from Quasar Luminosity Functions at z > 4: Implications for Faint and Low-mass Populations in JWST},} \apj, 969, 69, \dodoi{10.3847/1538-4357/ad46f9}

\bibitem[{Z. {Li} {et~al.}(2025){Li}, {Inayoshi}, {Chen}, {Ichikawa}, \& {Ho}}]{Li2025zhengrong}
{Li}, Z., {Inayoshi}, K., {Chen}, K., {Ichikawa}, K., \& {Ho}, L.~C. 2025, \bibinfo{title}{{Little Red Dots: Rapidly Growing Black Holes Reddened by Extended Dusty Flows},} \apj, 980, 36, \dodoi{10.3847/1538-4357/ada5fb}

\bibitem[{X. {Lin} {et~al.}(2024){Lin}, {Wang}, {Fan}, {Cai}, {Champagne}, {Sun}, {Volonteri}, {Yang}, {Hennawi}, {Ba{\~n}ados}, {Barth}, {Eilers}, {Farina}, {Liu}, {Jin}, {Jun}, {Lupi}, {Kakiichi}, {Mazzucchelli}, {Onoue}, {Pan}, {Pizzati}, {Rojas-Ruiz}, {Schindler}, {Trakhtenbrot}, {Shen}, {Trebitsch}, {Zhuang}, {Endsley}, {Meyer}, {Li}, {Li}, {Pudoka}, {Tee}, {Wu}, \& {Zhang}}]{Lin2024aspire}
{Lin}, X., {Wang}, F., {Fan}, X., {et~al.} 2024, \bibinfo{title}{{A SPectroscopic Survey of Biased Halos In the Reionization Era (ASPIRE): Broad-line AGN at z = 4‑5 Revealed by JWST/NIRCam WFSS},} \apj, 974, 147, \dodoi{10.3847/1538-4357/ad6565}

\bibitem[{X. {Lin} {et~al.}(2025){Lin}, {Fan}, {Cai}, {Bian}, {Liu}, {Sun}, {Ma}, {Greene}, {Strauss}, {Green}, {Lyu}, {Champagne}, {Goulding}, {Inayoshi}, {Jin}, {Leung}, {Li}, {Liu}, {Mao}, {Pudoka}, {Tee}, {Wang}, {Wang}, {Wu}, {Yang}, {Zhang}, \& {Zhu}}]{Lin2025local}
{Lin}, X., {Fan}, X., {Cai}, Z., {et~al.} 2025, \bibinfo{title}{{The Discovery of Little Red Dots in the Local Universe: Signatures of Cool Gas Envelopes},} arXiv e-prints, arXiv:2507.10659, \dodoi{10.48550/arXiv.2507.10659}

\bibitem[{H. {Liu} {et~al.}(2025){Liu}, {Jiang}, {Quataert}, {Greene}, \& {Ma}}]{Liu2025hanpu}
{Liu}, H., {Jiang}, Y.-F., {Quataert}, E., {Greene}, J.~E., \& {Ma}, Y. 2025, \bibinfo{title}{{The Balmer Break and Optical Continuum of Little Red Dots From Super-Eddington Accretion},} arXiv e-prints, arXiv:2507.07190, \dodoi{10.48550/arXiv.2507.07190}

\bibitem[{Y. {Ma} {et~al.}(2025{\natexlab{a}}){Ma}, {Greene}, {Setton}, {Volonteri}, {Leja}, {Wang}, {Bezanson}, {Brammer}, {Cutler}, {Dayal}, {van Dokkum}, {Furtak}, {Glazebrook}, {Goulding}, {de Graaff}, {Kokorev}, {Labbe}, {Pan}, {Price}, {Weaver}, {Williams}, {Whitaker}, \& {Zitrin}}]{Ma2025}
{Ma}, Y., {Greene}, J.~E., {Setton}, D.~J., {et~al.} 2025{\natexlab{a}}, \bibinfo{title}{{UNCOVER: 404 Error{\textemdash}Models Not Found for the Triply Imaged Little Red Dot A2744-QSO1},} \apj, 981, 191, \dodoi{10.3847/1538-4357/ada613}

\bibitem[{Y. {Ma} {et~al.}(2025{\natexlab{b}}){Ma}, {Greene}, {Setton}, {Goulding}, {Annunziatella}, {Fan}, {Kokorev}, {Labbe}, {Li}, {Lin}, {Marchesini}, {Matthee}, {Robbins}, {Sajina}, {Sawicki}, \& {Telford}}]{Ma2025z2lrd}
{Ma}, Y., {Greene}, J.~E., {Setton}, D.~J., {et~al.} 2025{\natexlab{b}}, \bibinfo{title}{{Counting Little Red Dots at $z<4$ with Ground-based Surveys and Spectroscopic Follow-up},} arXiv e-prints, arXiv:2504.08032, \dodoi{10.48550/arXiv.2504.08032}

\bibitem[{D. {Marchesini} {et~al.}(2009){Marchesini}, {van Dokkum}, {F{\"o}rster Schreiber}, {Franx}, {Labb{\'e}}, \& {Wuyts}}]{Marchesini2009smf}
{Marchesini}, D., {van Dokkum}, P.~G., {F{\"o}rster Schreiber}, N.~M., {et~al.} 2009, \bibinfo{title}{{The Evolution of the Stellar Mass Function of Galaxies from z = 4.0 and the First Comprehensive Analysis of its Uncertainties: Evidence for Mass-Dependent Evolution},} \apj, 701, 1765, \dodoi{10.1088/0004-637X/701/2/1765}

\bibitem[{R. {Massey} {et~al.}(2010){Massey}, {Stoughton}, {Leauthaud}, {Rhodes}, {Koekemoer}, {Ellis}, \& {Shaghoulian}}]{Massey2010cosmos_hst_reduction}
{Massey}, R., {Stoughton}, C., {Leauthaud}, A., {et~al.} 2010, \bibinfo{title}{{Pixel-based correction for Charge Transfer Inefficiency in the Hubble Space Telescope Advanced Camera for Surveys},} \mnras, 401, 371, \dodoi{10.1111/j.1365-2966.2009.15638.x}

\bibitem[{J. {Matthee} {et~al.}(2024){Matthee}, {Naidu}, {Brammer}, {Chisholm}, {Eilers}, {Goulding}, {Greene}, {Kashino}, {Labbe}, {Lilly}, {Mackenzie}, {Oesch}, {Weibel}, {Wuyts}, {Xiao}, {Bordoloi}, {Bouwens}, {van Dokkum}, {Illingworth}, {Kramarenko}, {Maseda}, {Mason}, {Meyer}, {Nelson}, {Reddy}, {Shivaei}, {Simcoe}, \& {Yue}}]{Matthee2024}
{Matthee}, J., {Naidu}, R.~P., {Brammer}, G., {et~al.} 2024, \bibinfo{title}{{Little Red Dots: An Abundant Population of Faint Active Galactic Nuclei at z {\ensuremath{\sim}} 5 Revealed by the EIGER and FRESCO JWST Surveys},} \apj, 963, 129, \dodoi{10.3847/1538-4357/ad2345}

\bibitem[{H.~J. {McCracken} {et~al.}(2012){McCracken}, {Milvang-Jensen}, {Dunlop}, {Franx}, {Fynbo}, {Le F{\`e}vre}, {Holt}, {Caputi}, {Goranova}, {Buitrago}, {Emerson}, {Freudling}, {Hudelot}, {L{\'o}pez-Sanjuan}, {Magnard}, {Mellier}, {M{\o}ller}, {Nilsson}, {Sutherland}, {Tasca}, \& {Zabl}}]{McCracken2012uvista}
{McCracken}, H.~J., {Milvang-Jensen}, B., {Dunlop}, J., {et~al.} 2012, \bibinfo{title}{{UltraVISTA: a new ultra-deep near-infrared survey in COSMOS},} \aap, 544, A156, \dodoi{10.1051/0004-6361/201219507}

\bibitem[{R.~P. {Naidu} {et~al.}(2025){Naidu}, {Matthee}, {Katz}, {de Graaff}, {Oesch}, {Smith}, {Greene}, {Brammer}, {Weibel}, {Hviding}, {Chisholm}, {Labb\textbackslash'e}, {Simcoe}, {Witten}, {Atek}, {Baggen}, {Belli}, {Bezanson}, {Boogaard}, {Bose}, {Covelo-Paz}, {Dayal}, {Fudamoto}, {Furtak}, {Giovinazzo}, {Goulding}, {Gronke}, {Heintz}, {Hirschmann}, {Illingworth}, {Inoue}, {Johnson}, {Leja}, {Leonova}, {McConachie}, {Maseda}, {Natarajan}, {Nelson}, {Setton}, {Shivaei}, {Sobral}, {Stefanon}, {Tacchella}, {Toft}, {Torralba}, {van Dokkum}, {van der Wel}, {Volonteri}, {Walter}, {Wang}, \& {Watson}}]{Naidu2025bhstar}
{Naidu}, R.~P., {Matthee}, J., {Katz}, H., {et~al.} 2025, \bibinfo{title}{{A ``Black Hole Star'' Reveals the Remarkable Gas-Enshrouded Hearts of the Little Red Dots},} arXiv e-prints, arXiv:2503.16596, \dodoi{10.48550/arXiv.2503.16596}

\bibitem[{M. {Niida} {et~al.}(2020){Niida}, {Nagao}, {Ikeda}, {Akiyama}, {Matsuoka}, {He}, {Matsuoka}, {Toba}, {Onoue}, {Kobayashi}, {Taniguchi}, {Furusawa}, {Harikane}, {Imanishi}, {Kashikawa}, {Kawaguchi}, {Komiyama}, {Shirakata}, {Terashima}, \& {Ueda}}]{Niida2020QLF}
{Niida}, M., {Nagao}, T., {Ikeda}, H., {et~al.} 2020, \bibinfo{title}{{The Faint End of the Quasar Luminosity Function at z {\ensuremath{\sim}} 5 from the Subaru Hyper Suprime-Cam Survey},} \apj, 904, 89, \dodoi{10.3847/1538-4357/abbe11}

\bibitem[{J.~B. {Oke} \& J.~E. {Gunn}(1983){Oke} \& {Gunn}}]{Oke&Gunn1983}
{Oke}, J.~B., \& {Gunn}, J.~E. 1983, \bibinfo{title}{{Secondary standard stars for absolute spectrophotometry.},} \apj, 266, 713, \dodoi{10.1086/160817}

\bibitem[{F. {Pacucci} \& R. {Narayan}(2024){Pacucci} \& {Narayan}}]{Pacucci&Narayan2024}
{Pacucci}, F., \& {Narayan}, R. 2024, \bibinfo{title}{{Mildly Super-Eddington Accretion onto Slowly Spinning Black Holes Explains the X-Ray Weakness of the Little Red Dots},} \apj, 976, 96, \dodoi{10.3847/1538-4357/ad84f7}

\bibitem[{I. {Pasha} \& T.~B. {Miller}(2023){Pasha} \& {Miller}}]{Pasha&Miller2023pysersic}
{Pasha}, I., \& {Miller}, T.~B. 2023, \bibinfo{title}{{pysersic: A Python package for determining galaxy structural properties via Bayesian inference, accelerated with jax},} The Journal of Open Source Software, 8, 5703, \dodoi{10.21105/joss.05703}

\bibitem[{G.~T. {Richards} {et~al.}(2006){Richards}, {Lacy}, {Storrie-Lombardi}, {Hall}, {Gallagher}, {Hines}, {Fan}, {Papovich}, {Vanden Berk}, {Trammell}, {Schneider}, {Vestergaard}, {York}, {Jester}, {Anderson}, {Budav{\'a}ri}, \& {Szalay}}]{Richards2006}
{Richards}, G.~T., {Lacy}, M., {Storrie-Lombardi}, L.~J., {et~al.} 2006, \bibinfo{title}{{Spectral Energy Distributions and Multiwavelength Selection of Type 1 Quasars},} \apjs, 166, 470, \dodoi{10.1086/506525}

\bibitem[{P. {Rinaldi} {et~al.}(2024){Rinaldi}, {Bonaventura}, {Rieke}, {Alberts}, {Caputi}, {Baker}, {Baum}, {Bhatawdekar}, {Bunker}, {Carniani}, {Curtis-Lake}, {D'Eugenio}, {Egami}, {Ji}, {Hainline}, {Helton}, {Lin}, {Lyu}, {Johnson}, {Ma}, {Maiolino}, {P{\'e}rez-Gonz{\'a}lez}, {Rieke}, {Robertson}, {Shivaei}, {Stone}, {Sun}, {Tacchella}, {{\"U}bler}, {Williams}, {Willmer}, {Willott}, {Zhang}, \& {Zhu}}]{Rinaldi2024}
{Rinaldi}, P., {Bonaventura}, N., {Rieke}, G.~H., {et~al.} 2024, \bibinfo{title}{{Not Just a Dot: the complex UV morphology and underlying properties of Little Red Dots},} arXiv e-prints, arXiv:2411.14383, \dodoi{10.48550/arXiv.2411.14383}

\bibitem[{P. {Rinaldi} {et~al.}(2025){Rinaldi}, {Rieke}, {Wu}, {Gilbert}, {Pacucci}, {Barchiesi}, {Alberts}, {Carniani}, {Bunker}, {Bhatawdekar}, {D'Eugenio}, {Ji}, {Johnson}, {Hainline}, {Kokorev}, {Kumari}, {Iani}, {Lyu}, {Maiolino}, {Parlanti}, {Robertson}, {Sun}, {Vignali}, {Williams}, {Willmer}, \& {Zhu}}]{Rinaldi2025}
{Rinaldi}, P., {Rieke}, G.~H., {Wu}, Z., {et~al.} 2025, \bibinfo{title}{{Beyond the Dot: an LRD-like nucleus at the Heart of an IR-Bright Galaxy and its implications for high-redshift LRDs},} arXiv e-prints, arXiv:2507.17738, \dodoi{10.48550/arXiv.2507.17738}

\bibitem[{V. {Rusakov} {et~al.}(2025){Rusakov}, {Watson}, {Nikopoulos}, {Brammer}, {Gottumukkala}, {Harvey}, {Heintz}, {Nielsen}, {Sim}, {Sneppen}, {Vijayan}, {Adams}, {Austin}, {Conselice}, {Goolsby}, {Toft}, \& {Witstok}}]{Rusakov2025}
{Rusakov}, V., {Watson}, D., {Nikopoulos}, G.~P., {et~al.} 2025, \bibinfo{title}{{JWST's little red dots: an emerging population of young, low-mass AGN cocooned in dense ionized gas},} arXiv e-prints, arXiv:2503.16595, \dodoi{10.48550/arXiv.2503.16595}

\bibitem[{M. {Sawicki} {et~al.}(2019){Sawicki}, {Arnouts}, {Huang}, {Coupon}, {Golob}, {Gwyn}, {Foucaud}, {Moutard}, {Iwata}, {Liu}, {Chen}, {Desprez}, {Harikane}, {Ono}, {Strauss}, {Tanaka}, {Thibert}, {Balogh}, {Bundy}, {Chapman}, {Gunn}, {Hsieh}, {Ilbert}, {Jing}, {LeF{\`e}vre}, {Li}, {Matsuda}, {Miyazaki}, {Nagao}, {Nishizawa}, {Ouchi}, {Shimasaku}, {Silverman}, {de la Torre}, {Tresse}, {Wang}, {Willott}, {Yamada}, {Yang}, \& {Yee}}]{Sawicki2019clauds}
{Sawicki}, M., {Arnouts}, S., {Huang}, J., {et~al.} 2019, \bibinfo{title}{{The CFHT large area U-band deep survey (CLAUDS)},} \mnras, 489, 5202, \dodoi{10.1093/mnras/stz2522}

\bibitem[{P. {Schechter}(1976){Schechter}}]{Schechter1976}
{Schechter}, P. 1976, \bibinfo{title}{{An analytic expression for the luminosity function for galaxies.},} \apj, 203, 297, \dodoi{10.1086/154079}

\bibitem[{M. {Schmidt}(1968){Schmidt}}]{Schmidt1968}
{Schmidt}, M. 1968, \bibinfo{title}{{Space Distribution and Luminosity Functions of Quasi-Stellar Radio Sources},} \apj, 151, 393, \dodoi{10.1086/149446}

\bibitem[{N. {Scoville} {et~al.}(2007){Scoville}, {Aussel}, {Brusa}, {Capak}, {Carollo}, {Elvis}, {Giavalisco}, {Guzzo}, {Hasinger}, {Impey}, {Kneib}, {LeFevre}, {Lilly}, {Mobasher}, {Renzini}, {Rich}, {Sanders}, {Schinnerer}, {Schminovich}, {Shopbell}, {Taniguchi}, \& {Tyson}}]{Scoville2007cosmos}
{Scoville}, N., {Aussel}, H., {Brusa}, M., {et~al.} 2007, \bibinfo{title}{{The Cosmic Evolution Survey (COSMOS): Overview},} \apjs, 172, 1, \dodoi{10.1086/516585}

\bibitem[{J.~L. {S{\'e}rsic}(1963){S{\'e}rsic}}]{Sersic1963}
{S{\'e}rsic}, J.~L. 1963, \bibinfo{title}{{Influence of the atmospheric and instrumental dispersion on the brightness distribution in a galaxy},} Boletin de la Asociacion Argentina de Astronomia La Plata Argentina, 6, 41

\bibitem[{D.~J. {Setton} {et~al.}(2024){Setton}, {Greene}, {de Graaff}, {Ma}, {Leja}, {Matthee}, {Bezanson}, {Boogaard}, {Cleri}, {Katz}, {Labbe}, {Maseda}, {McConachie}, {Miller}, {Price}, {Suess}, {van Dokkum}, {Wang}, {Weibel}, {Whitaker}, \& {Williams}}]{Setton2024inflection}
{Setton}, D.~J., {Greene}, J.~E., {de Graaff}, A., {et~al.} 2024, \bibinfo{title}{{Little Red Dots at an Inflection Point: Ubiquitous ``V-Shaped'' Turnover Consistently Occurs at the Balmer Limit},} arXiv e-prints, arXiv:2411.03424, \dodoi{10.48550/arXiv.2411.03424}

\bibitem[{D.~J. {Setton} {et~al.}(2025){Setton}, {Greene}, {Spilker}, {Williams}, {Labbe}, {Ma}, {Wang}, {Whitaker}, {Leja}, {de Graaff}, {Alberts}, {Bezanson}, {Boogaard}, {Brammer}, {Cutler}, {Cleri}, {Cooper}, {Dayal}, {Fujimoto}, {Furtak}, {Goulding}, {Hirschmann}, {Kokorev}, {Maseda}, {McConachie}, {Matthee}, {Miller}, {Naidu}, {Oesch}, {Pan}, {Price}, {Suess}, {Weaver}, {Xiao}, {Zhang}, \& {Zitrin}}]{Setton2025alma}
{Setton}, D.~J., {Greene}, J.~E., {Spilker}, J.~S., {et~al.} 2025, \bibinfo{title}{{A confirmed deficit of hot and cold dust emission in the most luminous Little Red Dots},} arXiv e-prints, arXiv:2503.02059, \dodoi{10.48550/arXiv.2503.02059}

\bibitem[{A. {S{\k{a}}dowski} {et~al.}(2015){S{\k{a}}dowski}, {Narayan}, {Tchekhovskoy}, {Abarca}, {Zhu}, \& {McKinney}}]{Sadowski2015mdotL}
{S{\k{a}}dowski}, A., {Narayan}, R., {Tchekhovskoy}, A., {et~al.} 2015, \bibinfo{title}{{Global simulations of axisymmetric radiative black hole accretion discs in general relativity with a mean-field magnetic dynamo},} \mnras, 447, 49, \dodoi{10.1093/mnras/stu2387}

\bibitem[{T.~S. {Tanaka} {et~al.}(2025){Tanaka}, {Akins}, {Harikane}, {Silverman}, {Casey}, {Inayoshi}, {Schindler}, {Shimasaku}, {Kocevski}, {Onoue}, {Faisst}, {Robertson}, {Kokorev}, {Shuntov}, {Koekemoer}, {Franco}, {Liu}, {Taylor}, {Kartaltepe}, {Bosman}, {Champagne}, {Kakiichi}, {Zhang}, {Newman}, {Kakkad}, {Fei}, {Fujimoto}, {Li}, {Finkelstein}, {Li}, {Lambrides}, {Sommovigo}, {Zavala}, {Ito}, {Liu}, {Treister}, {Aravena}, {Gozaliasl}, {Zhang}, {Hatamnia}, {Umeda}, {Inoue}, {Yang}, {Ando}, {Arita}, {Ding}, {Matsui}, {Shibanuma}, {Magdis}, {Zhuang}, {Fan}, {Li}, {Liu}, {Lyu}, {Rhodes}, {Toft}, {Wang}, {Zou}, {Arango-Toro}, {Battisti}, {Gillman}, {Khostovan}, \& {Long}}]{Tanaka2025z10lrd}
{Tanaka}, T.~S., {Akins}, H.~B., {Harikane}, Y., {et~al.} 2025, \bibinfo{title}{{Discovery of a Little Red Dot candidate at $zrsim10$ in COSMOS-Web based on MIRI-NIRCam selection},} arXiv e-prints, arXiv:2508.00057, \dodoi{10.48550/arXiv.2508.00057}

\bibitem[{M.~J. {Temple} {et~al.}(2021){Temple}, {Hewett}, \& {Banerji}}]{Temple2021}
{Temple}, M.~J., {Hewett}, P.~C., \& {Banerji}, M. 2021, \bibinfo{title}{{Modelling type 1 quasar colours in the era of Rubin and Euclid},} \mnras, 508, 737, \dodoi{10.1093/mnras/stab2586}

\bibitem[{A. {Torralba} {et~al.}(2025){Torralba}, {Matthee}, {Pezzulli}, {Urrutia}, {Gronke}, {Mascia}, {D'Eugenio}, {Di Cesare}, {Eilers}, {Greene}, {Iani}, {Ishikawa}, {Mackenzie}, {Naidu}, {Navarrete}, \& {Kotiwale}}]{Torralba2025monster_muse}
{Torralba}, A., {Matthee}, J., {Pezzulli}, G., {et~al.} 2025, \bibinfo{title}{{A weak Ly$α$ halo for an extremely bright Little Red Dot. Indications of enshrouded SMBH growth},} arXiv e-prints, arXiv:2505.09542, \dodoi{10.48550/arXiv.2505.09542}

\bibitem[{A. {Trinca} {et~al.}(2024){Trinca}, {Valiante}, {Schneider}, {Juod{\v{z}}balis}, {Maiolino}, {Graziani}, {Lupi}, {Natarajan}, {Volonteri}, \& {Zana}}]{Trinca2024}
{Trinca}, A., {Valiante}, R., {Schneider}, R., {et~al.} 2024, \bibinfo{title}{{Episodic super-Eddington accretion as a clue to Overmassive Black Holes in the early Universe},} arXiv e-prints, arXiv:2412.14248, \dodoi{10.48550/arXiv.2412.14248}

\bibitem[{M. {Volonteri} {et~al.}(2017){Volonteri}, {Reines}, {Atek}, {Stark}, \& {Trebitsch}}]{Volonteri2017}
{Volonteri}, M., {Reines}, A.~E., {Atek}, H., {Stark}, D.~P., \& {Trebitsch}, M. 2017, \bibinfo{title}{{High-redshift Galaxies and Black Holes Detectable with the JWST: A Population Synthesis Model from Infrared to X-Rays},} \apj, 849, 155, \dodoi{10.3847/1538-4357/aa93f1}

\bibitem[{B. {Wang} {et~al.}(2025){Wang}, {de Graaff}, {Davies}, {Greene}, {Leja}, {Brammer}, {Goulding}, {Miller}, {Suess}, {Weibel}, {Williams}, {Bezanson}, {Boogaard}, {Cleri}, {Hirschmann}, {Katz}, {Labb{\'e}}, {Maseda}, {Matthee}, {McConachie}, {Naidu}, {Oesch}, {Rix}, {Setton}, \& {Whitaker}}]{Wang2025brd}
{Wang}, B., {de Graaff}, A., {Davies}, R.~L., {et~al.} 2025, \bibinfo{title}{{RUBIES: JWST/NIRSpec Confirmation of an Infrared-luminous, Broad-line Little Red Dot with an Ionized Outflow},} \apj, 984, 121, \dodoi{10.3847/1538-4357/adc1ca}

\bibitem[{K.-y. {Watarai} {et~al.}(2000){Watarai}, {Fukue}, {Takeuchi}, \& {Mineshige}}]{Watarai2000mdotL}
{Watarai}, K.-y., {Fukue}, J., {Takeuchi}, M., \& {Mineshige}, S. 2000, \bibinfo{title}{{Galactic Black-Hole Candidates Shining at the Eddington Luminosity},} \pasj, 52, 133, \dodoi{10.1093/pasj/52.1.133}

\bibitem[{C.~C. {Williams} {et~al.}(2024){Williams}, {Alberts}, {Ji}, {Hainline}, {Lyu}, {Rieke}, {Endsley}, {Suess}, {Sun}, {Johnson}, {Florian}, {Shivaei}, {Rujopakarn}, {Baker}, {Bhatawdekar}, {Boyett}, {Bunker}, {Cameron}, {Carniani}, {Charlot}, {Curtis-Lake}, {DeCoursey}, {de Graaff}, {Egami}, {Eisenstein}, {Gibson}, {Hausen}, {Helton}, {Maiolino}, {Maseda}, {Nelson}, {P{\'e}rez-Gonz{\'a}lez}, {Rieke}, {Robertson}, {Saxena}, {Tacchella}, {Willmer}, \& {Willott}}]{Williams2024}
{Williams}, C.~C., {Alberts}, S., {Ji}, Z., {et~al.} 2024, \bibinfo{title}{{The Galaxies Missed by Hubble and ALMA: The Contribution of Extremely Red Galaxies to the Cosmic Census at 3 < z < 8},} \apj, 968, 34, \dodoi{10.3847/1538-4357/ad3f17}

\bibitem[{M. {Xiao} {et~al.}(2025){Xiao}, {Oesch}, {Bing}, {Elbaz}, {Matthee}, {Fudamoto}, {Fujimoto}, {Marques-Chaves}, {Williams}, {Dessauges-Zavadsky}, {Valentino}, {Brammer}, {Covelo-Paz}, {Daddi}, {Fynbo}, {Gillman}, {Ginolfi}, {Giovinazzo}, {Greene}, {Gu}, {Illingworth}, {Inayoshi}, {Kokorev}, {Meyer}, {Naidu}, {Reddy}, {Schaerer}, {Shapley}, {Stefanon}, {Steinhardt}, {Setton}, {Vestergaard}, \& {Wang}}]{Xiao2025noema}
{Xiao}, M., {Oesch}, P.~A., {Bing}, L., {et~al.} 2025, \bibinfo{title}{{No [CII] or dust detection in two Little Red Dots at z$_{\rm spec}$ > 7},} arXiv e-prints, arXiv:2503.01945, \dodoi{10.48550/arXiv.2503.01945}

\bibitem[{M. {Yue} {et~al.}(2024){Yue}, {Eilers}, {Ananna}, {Panagiotou}, {Kara}, \& {Miyaji}}]{Yue2024}
{Yue}, M., {Eilers}, A.-C., {Ananna}, T.~T., {et~al.} 2024, \bibinfo{title}{{Stacking X-Ray Observations of ``Little Red Dots'': Implications for Their Active Galactic Nucleus Properties},} \apjl, 974, L26, \dodoi{10.3847/2041-8213/ad7eba}

\bibitem[{M.-Y. {Zhuang} {et~al.}(2025){Zhuang}, {Li}, {Shen}, {Lin}, {Shapley}, {Wang}, {Wu}, \& {Yang}}]{Zhuang2025nexusLRD}
{Zhuang}, M.-Y., {Li}, J., {Shen}, Y., {et~al.} 2025, \bibinfo{title}{{NEXUS: A Spectroscopic Census of Broad-line AGNs and Little Red Dots at $3\lesssim z\lesssim 6$},} arXiv e-prints, arXiv:2505.20393, \dodoi{10.48550/arXiv.2505.20393}

\end{thebibliography}
\bibliographystyle{aasjournal}

\end{CJK*}
\end{document}